\documentclass[11pt,a4paper]{article}
\pdfoutput=1

\usepackage{jheppub}

\usepackage[utf8]{inputenc}
\usepackage[T1]{fontenc}
\usepackage{afterpage}
\usepackage{rotating}

\newcommand{\Dmq}{\Delta m^2}
\newcommand{\Eps}{\varepsilon}
\newcommand{\Qwq}{\mathcal{Q}^2}
\makeatletter
\DeclareRobustCommand\recite[1]{\begingroup\@fileswfalse\cite{#1}\endgroup}
\makeatother

\graphicspath{{Figures/}}

\allowdisplaybreaks

\preprint{
  \begin{flushright}
    YITP-SB-19-38\\
    IFT-UAM/CSIC-19-152\\
    IFIC-19-49
  \end{flushright}
}

\title{Improved global fit to Non-Standard neutrino Interactions using
  COHERENT energy and timing data}

\author[a]{Pilar Coloma,}
\affiliation[a]{Instituto de Física Corpuscular, Universitat de
  València and CSIC, Edificio Institutos de Investigación, Calle
  Catedrático José Beltrán 2, E-46980 Valencia, Spain}
\emailAdd{pcoloma@ific.uv.es}

\author[b]{Ivan Esteban,}
\affiliation[b]{Departament de Física Quàntica i Astrofísica and
  Institut de Ciències del Cosmos, Universitat de Barcelona, Diagonal
  647, E-08028 Barcelona, Spain}
\emailAdd{ivan.esteban@fqa.ub.edu}

\author[b,c,d]{M.~C.~Gonzalez-Garcia,}
\affiliation[c]{Institució Catalana de Recerca i Estudis Avançats
  (ICREA), Pg.\ Lluis Companys 23, E-08010 Barcelona, Spain}
\affiliation[d]{C.N.~Yang Institute for Theoretical Physics, Stony
  Brook University, Stony Brook, NY 11794-3840, USA}
\emailAdd{maria.gonzalez-garcia@stonybrook.edu}

\author[e]{Michele Maltoni}
\affiliation[e]{Instituto de Física Teórica UAM/CSIC, Calle de
  Nicolás Cabrera 13--15, Universidad Autónoma de Madrid,
  Cantoblanco, E-28049 Madrid, Spain}
\emailAdd{michele.maltoni@csic.es}

\abstract{We perform a global fit to neutrino oscillation and coherent
  neutrino-nucleus scattering data, using both timing and energy
  information from the COHERENT experiment.  The results are used to
  set model-independent bounds on four-fermion effective operators
  inducing non-standard neutral-current neutrino interactions. We
  quantify the allowed ranges for their Wilson coefficients, as well
  as the status of the LMA-D solution, for a wide class of new physics
  models with arbitrary ratios between the strength of the operators
  involving up and down quarks. Our results are presented for the
  COHERENT experiment alone, as well as in combination with the global
  data from oscillation experiments.  We also quantify the dependence
  of our results for COHERENT with respect to the choice of quenching
  factor, nuclear form factor, and the treatment of the backgrounds.
}

\keywords{Neutrino physics, non-standard neutrino interactions,
  coherent elastic neutrino-nucleus scattering, neutrino oscillations}

\begin{document}

\maketitle

\section{Introduction}

Experiments measuring the flavor composition of neutrinos produced in
the Sun, in the Earth's atmosphere, in nuclear reactors and in
particle accelerators have established that lepton flavor is not
conserved in neutrino propagation, but it oscillates with a wavelength
which depends on distance and energy. This demonstrates beyond doubt
that neutrinos are massive and that the mass states are non-trivial
admixtures of flavor states~\cite{Pontecorvo:1967fh, Gribov:1968kq},
see Ref.~\cite{GonzalezGarcia:2007ib} for an overview.

Under the assumption that the Standard Model (SM) is the low-energy
effective model of a complete high-energy theory, a completely
model-independent parametrization of the possible effects of New
Physics (NP) at low energies is through the addition to the SM
Lagrangian of higher-dimensional operators which respect the SM gauge
group. The only dimension-five operator that can be built using just
SM fields is the Weinberg operator~\cite{Weinberg:1979sa}, which
coincidentally gives rise to neutrino masses. Given that the
observation of neutrino masses is one of the strongest indications of
physics beyond the SM, one may therefore expect additional effects
from higher dimensional operators. At dimension six, the allowed set
is larger and includes four-fermion operators affecting neutrino
production, propagation and detection processes, usually referred to
as Non-Standard neutrino Interactions (NSI).  For example, the
effective Lagrangian
\begin{equation}
  \label{eq:nsi-cc}
  \mathcal{L}_\text{NSI,CC} = \sum_{f,f',\alpha,\beta}
  2\sqrt{2} G_F \Eps_{\alpha\beta}^{ff',P}
  (\bar\nu_\alpha \gamma_\mu P_L \ell_\beta)
  (\bar f' \gamma^\mu P f )
  + \text{h.c.}
\end{equation}
would induce non-standard charged-current (CC) production and
detection mechanisms for neutrinos of flavor $\alpha$, while
\begin{equation}
  \label{eq:nsi-nc}
  \mathcal{L}_\text{NSI,NC} = \sum_{f,\alpha,\beta}
  2\sqrt{2} G_F \Eps_{\alpha \beta}^{f,P}
  (\bar\nu_\alpha \gamma_\mu P_L \nu_\beta)
  (\bar f \gamma^\mu P f)
  + \text{h.c.}
\end{equation}
would lead to new neutral-current (NC) interactions with the rest of
the SM fermions.  In both Eqs.~\eqref{eq:nsi-cc}
and~\eqref{eq:nsi-nc}, $f$ and $f'$ refer to SM fermions, $\ell$
denotes a SM charged lepton and $P$ can be either a left-handed or a
right-handed projection operator ($P_L$ or $P_R$, respectively). Note
that the new interactions may induce lepton flavor-changing processes
(if $\alpha \neq \beta$), or may lead to a modified interaction rate
with respect to the SM result (if $\alpha = \beta$).

While CC NSI are severely constrained from the precise measurement of
meson and muon decays (see for example Refs.~\cite{Davidson:2003ha,
  Biggio:2009nt, Biggio:2009kv}), constraining NC NSI is a much more
daunting task due to the technical challenges in the computation of
neutrino-nucleus interactions, and to the experimental difficulties
related to the measurement of neutrino NC interactions. In this case,
one may expect to see an observable effect in neutrino oscillations,
without entering in conflict with other experimental constraints.  Of
course, if the set of $d=6$ operators is obtained from a NP model at
high energies, electroweak gauge invariance generically implies that
the NC NSI operators can only be generated together with similar
operators involving charged leptons, for which the experimental
constraints are much tighter~\cite{Gavela:2008ra, Antusch:2008tz}.
However, recently it has been argued that viable NP models with light
mediators (\textit{i.e.}, below the electroweak scale) may lead to
large NC NSI effects which would affect neutrino oscillation
experiments without spoiling the precise determination of charged
lepton observables~\cite{Farzan:2017xzy, Miranda:2015dra,
  Farzan:2015doa, Farzan:2015hkd, Babu:2017olk,Denton:2018xmq}. For a
recent review on viable NSI models from light mediators see,
\textit{e.g.}, Ref.~\cite{Dev:2019anc}.\footnote{An alternative
  possibility would be to generate the NSI in radiative mass models,
  for example as in Ref.~\cite{Babu:2019mfe}.}

In this case, the best model-independent constraints available in the
literature for the vector operators inducing NC NSI come from global
fits to oscillation data, which are very sensitive to modifications in
the effective matter potential~\cite{Wolfenstein:1977ue,
  Mikheev:1986gs} felt by neutrinos as they propagate in a medium.
Since such modifications arise from a coherent effect, oscillation
bounds apply even to NSI induced by an ultra light mediators, as long
as their mass is $M_\text{med} \gtrsim 1/R_\text{Earth} \sim
\mathcal{O}(10^{-12})$~eV~\cite{GonzalezGarcia:2006vp}.  In
particular, oscillation experiments are sensitive to the combinations
of Wilson coefficients
\begin{equation}
  \Eps_{\alpha\beta}^f \equiv
  \Eps_{\alpha\beta}^{f,L} + \Eps_{\alpha\beta}^{f,R} \,,
\end{equation}
that is, to vector NSI.  It should be noted that, while oscillation
data are sensitive to all flavor-changing NSI, only the
\emph{differences} between flavor-diagonal NSI parameters induce
observable changes in the matter potential.  Consequently, oscillation
experiments can only bound five combinations of vector NSI: two
diagonal $\Eps_{\alpha\alpha}^f - \Eps_{\beta\beta}^f$, and three
non-diagonal $\Eps_{\alpha\beta}^f$ with $\alpha\neq\beta$.
Furthermore, in the presence of NSI~\cite{Wolfenstein:1977ue,
  Valle:1987gv, Guzzo:1991hi} a degeneracy exists in oscillation data,
leading to a qualitative change of the lepton mixing pattern. This was
first observed in the context of solar neutrinos, where for suitable
NSI the data can be explained by a mixing angle $\theta_{12}$ in the
second octant, the so-called LMA-Dark (LMA-D)~\cite{Miranda:2004nb}
solution. This is in sharp contrast to the established standard MSW
solution~\cite{Wolfenstein:1977ue, Mikheev:1986gs}, which requires a
mixing angle $\theta_{12}$ in the first octant.

The origin of the LMA-D solution is a degeneracy in the oscillation
probabilities due to a symmetry of the Hamiltonian describing neutrino
evolution in the presence of NSI~\cite{GonzalezGarcia:2011my,
  Gonzalez-Garcia:2013usa, Bakhti:2014pva, Coloma:2016gei}. For
neutrino oscillations in vacuum, it can be easily shown that a
simultaneous change in the neutrino mass ordering (that is, the sign
of $\Dmq_{31} \equiv m_3^2 - m_1^2$, with $m_i$ being the masses of
the three neutrino mass eigenstates), the octant of the solar mixing
angle $\theta_{12}$ and a shift in the leptonic CP-violating phase
$\delta_\text{CP}$ leaves the oscillatory pattern of neutrinos
completely unaffected~\cite{Bakhti:2014pva, Coloma:2016gei}.  Although
this degeneracy is broken in presence of a standard matter potential,
this is no longer the case if NSI are
allowed~\cite{GonzalezGarcia:2011my, Gonzalez-Garcia:2013usa,
  Coloma:2016gei}.  Hence, the LMA-D degeneracy makes it impossible to
determine the neutrino mass ordering by oscillation experiments
alone~\cite{Coloma:2016gei}, and therefore jeopardizes one of the main
goals of the upcoming neutrino oscillation program.

For oscillation experiments performed in matter with a uniform
neutron/proton ratio the LMA-D degeneracy stands as exact.  Possible
ways to lift it are through the combination of oscillation data
obtained in environments with different chemical compositions, or for
experiments observing neutrino oscillations in a matter potential
whose neutron/proton ratio changes sizeably along the neutrino path
(as in the case neutrino propagation inside the Sun).  This also
implies that the results depend on the relative strength of the
couplings to up and down quarks.  However, global fits to present data
show that the LMA-D solution is still pretty much allowed by
oscillation data alone~\cite{Gonzalez-Garcia:2013usa,
  Esteban:2018ppq}.

As discussed in Refs.~\cite{Coloma:2017ncl, Coloma:2017egw,
  Coloma:2016gei, Escrihuela:2009up, Miranda:2004nb}, non-oscillation
data (such as those collected by neutrino scattering experiments) is
needed to break this degeneracy.  However, for light enough mediators
($M_\text{med} \lesssim \mathcal{O}(10)$~GeV) bounds from deep
inelastic neutrino scattering experiments, such as
CHARM~\cite{Dorenbosch:1986tb} and NuTeV~\cite{Zeller:2001hh}, can be
successfully avoided~\cite{Farzan:2015doa, Coloma:2016gei,
  Coloma:2017egw}.  Thus, if NSI is generated by mediators as light as
about 10~MeV the degeneracy can only be broken through the combination
with results on coherent neutrino-nucleus
scattering~\cite{Freedman:1973yd} (CE$\nu$NS).  In fact, CE$\nu$NS has
been recently observed for the first time by the COHERENT
experiment~\cite{Akimov:2017ade} using neutrinos produced at the
Spallation Neutron Source (SNS) sited at the Oak Ridge National
Laboratory. The analysis of COHERENT data allows to constrain two of
the three flavor-diagonal NSI operators, since the neutrino flux
contains both muon and electron neutrinos. In combination with the
oscillation analysis, this allows for the independent determination of
all the three flavor-diagonal coefficients and has a strong impact on
the LMA-D degeneracy.

To this end, in Ref.~\cite{Coloma:2017ncl} we combined the first
results from the COHERENT experiment~\cite{Akimov:2017ade} with
previous bounds on NSI from a global fit to oscillation
data~\cite{Gonzalez-Garcia:2013usa}. The outcome of such analysis
showed that it was already possible to reject the LMA-D degeneracy
beyond $3\sigma$ in models where the NSI operators involved only a
single quark flavor (either up or down).  In addition, the study
presented in Ref.~\cite{Coloma:2017ncl} also yielded the first
independent determination of the three flavor diagonal NSI couplings
to up or down quarks. Subsequently, in Ref.~\cite{Esteban:2018ppq} we
updated our analysis of oscillation experiments to account for the
most recent data which had become available, and we also generalized
our framework to account for a NSI operators with both up and down
quarks at the same time, under the restriction that the neutrino
flavor structure of the NSI interactions is independent of the quark
type. More explicitly, in Ref.~\cite{Esteban:2018ppq} we assumed that
the Wilson coefficients for the vector couplings to quarks could be
parametrized in terms of a quark-indepenent matrix
$\Eps_{\alpha\beta}^\eta$ and an angle $\eta$ as
\begin{equation}
  \label{eq:eps-fact}
  \Eps_{\alpha\beta}^u = \frac{\sqrt{5}}{3}
  (2 \cos\eta - \sin\eta)\, \Eps_{\alpha\beta}^\eta
  \quad\text{and}\quad
  \Eps_{\alpha\beta}^d = \frac{\sqrt{5}}{3}
  (2 \sin\eta - \cos\eta)\, \Eps_{\alpha\beta}^\eta \,.
\end{equation}
The analysis in Ref.~\cite{Esteban:2018ppq} concluded that the LMA-D
degeneracy is a common feature of the results of the oscillation
analysis for a wide range of values of $\eta$, \textit{i.e.}, for a
broad spectrum of up-to-down NSI coupling strengths. It also showed
that the COHERENT data from Ref.~\cite{Akimov:2017ade} was able to
lift the degeneracy for a large subset of NSI models.

More recently the COHERENT collaboration released the detailed timing
and energy information~\cite{Akimov:2018vzs} of their data. This
allows for further tests of NP and it has been used to this effect in
a series of recent works~\cite{Papoulias:2019xaw, Han:2019zkz,
  Giunti:2019xpr, Cadeddu:2019eta, Khan:2019cvi, Miranda:2019wdy,
  Dutta:2019eml, Dutta:2019nbn, Papoulias:2019lfi, Huang:2019ene,
  Cadeddu:2018dux}.  With this motivation, in this work we quantify
the effect of including the timing and energy information of COHERENT
data on the bounds on NC NSI and, in particular, on the global
analysis in combination with oscillation data. In doing so we consider
the following novel aspects in the analysis of COHERENT energy and
timing data with respect to the analyses performed by other
groups~\cite{Papoulias:2019xaw, Han:2019zkz, Giunti:2019xpr,
  Cadeddu:2019eta, Khan:2019cvi, Miranda:2019wdy, Dutta:2019eml,
  Dutta:2019nbn, Papoulias:2019lfi, Huang:2019ene, Cadeddu:2018dux}:
\begin{itemize}
\item A broader range of NSI models, with couplings parametrized in
  Eq.~\eqref{eq:eps-fact}, has been considered.

\item Following recent discussions in the community regarding the
  validity of the quenching factor (QF) assumed by the experimental
  collaboration, in our simulation of the COHERENT signal we use a
  variety of QF implementations. In particular we introduce a new
  parametrization from our own fit to the calibration measurements
  performed by the TUNL group~\cite{Akimov:2017ade, Akimov:2018vzs},
  never considered before in the literature. The impact of the QF
  parametrization on the fit is therefore quantified and clarified.

\item We study the effect of the nuclear form factor employed and in
  doing so we use, for the first time, the results of a state-of-the
  art theoretical calculation of the nuclear form factor provided to
  us by the authors of Refs.~\cite{menendez, Klos:2013rwa}.

\item Quantitatively, the most relevant novelty in our analysis is our
  reevaluation of the steady-state background, as opposed to the
  ad-hoc parametrization used in the experimental data release (and
  employed in one way or another in the subsequent phenomenological
  analyses by different groups). As we will show, using that
  parametrization results into the disfavouring of the Standard Model
  at the $2\sigma$ level. We show how this is not the case when we use
  their own beam-off data to determine their steady-state background.
\end{itemize}
In addition, as mentioned above, we present for the first time the
results of combining the information from the COHERENT time and energy
spectrum with that from the analysis of oscillation data.  We quantify
the improvement of the bounds on the NSI coefficients by comparing
with the results of Ref.~\cite{Esteban:2018ppq}. Our results show that
the LMA-D solution is more significantly disfavoured for a wider range
of NSI models. Furthermore, our scrutiny of the possible variations of
the COHERENT time and energy information (described above), puts the
final bounds on more solid ground.

This paper is organized as follows. We start by briefly summarizing in
Sec.~\ref{sec:coh1} the framework for the evaluation of the CE$\nu$NS
event predictions at the SNS in the presence of
NSI. Section~\ref{sec:coh2} describes in detail our implementation of
the predicted timing and energy dependence of the event rates in
COHERENT, and the statistical analysis used.  Our results are then
presented in Sec.~\ref{sec:results}, first using only COHERENT data
(Sec.~\ref{sec:rescoh}) and then in combination with the results from
oscillation experiments (Sec.~\ref{sec:resglob}). In doing so we
discuss the dependence of the results on variations of the COHERENT
analysis associated to the choice of quenching factor, the nuclear
form factor, and the treatment of the backgrounds.  Finally in
Sec.~\ref{sec:summary} we summarize and draw our conclusions.

\section{Coherent elastic neutrino-nucleus scattering}
\label{sec:coh1}

At the SNS, an abundant flux of both $\pi^+$ and $\pi^-$ is produced
in proton-nucleus collisions in a mercury target. While the $\pi^-$
are absorbed by nuclei before they can decay, the $\pi^+$ lose energy
as they propagate and eventually decay at rest into $\pi^+ \to \mu^+
\nu_\mu$, followed by $\mu^+ \to e^+ \nu_e \bar\nu_\mu$. Since the
muon lifetime is much longer than that of the pion, the $\nu_\mu$
component is usually referred to as the prompt contribution to the
flux, as opposed to the delayed contributions from $\mu^+$ decay
($\bar\nu_\mu$ and $\nu_e$).

Given that the prompt neutrinos are a by-product of two-body decays at
rest, their contribution to the total flux is a monochromatic line at
$E_\text{pr} = (m_\pi^2 - m_\mu^2)/(2 m_\pi) \simeq 29.7$~MeV, where
$m_\pi$ and $m_\mu$ refer to the pion and muon masses,
respectively. Conversely, the delayed neutrino fluxes follow a
continuous spectra at energies $E_{\nu_e, \bar\nu_\mu} < m_\mu/2\simeq
52.8$~MeV. At a distance $\ell$ from the source, they read:
\begin{equation}
  \label{eq:COHflux}
  \begin{aligned}
    \frac{d\phi_{\nu_\mu}}{dE_\nu}
    &= \frac{1}{4\pi \ell^2 } \delta(E_\nu - E_\text{pr}) \,,
    \\
    \frac{d\phi_{\bar\nu_\mu}}{dE_\nu}
    &= \frac{1}{4\pi \ell^2} \frac{64}{m_\mu}
    \left[ \left( \frac{E_\nu}{m_\mu} \right)^2
      \left( \frac{3}{4} - \frac{E_\nu}{m_\mu} \right) \right],
    \\
    \frac{d\phi_{\nu_e}}{dE_\nu}
    &= \frac{1}{4\pi \ell^2} \frac{192}{m_\mu}
    \left[ \left( \frac{E_\nu}{m_\mu} \right)^2
      \left( \frac{1}{2} - \frac{E_\nu}{m_\mu} \right) \right],
  \end{aligned}
\end{equation}
and are normalized to each proton collision on the target. For
reference the distance $\ell$ at COHERENT is 19.3~m.

The differential cross section for coherent elastic neutrino-nucleus
scattering, for a neutrino with incident energy $E_\nu$ interacting
with a nucleus with $Z$ protons and $N$ neutrons,
reads~\cite{Freedman:1973yd}:
\begin{equation}
  \label{eq:xsec-SM}
  \frac{d\sigma_\text{SM} (T, E_\nu)}{dT}
  = \frac{G_F^2}{ 2 \pi}
  \Qwq (Z, N)  F^2(Q^2) M
  \left(2 - \frac{M T}{E_\nu^2} \right)
\end{equation}
where $G_F$ is the Fermi constant and $\Qwq$ is the weak charge of the
nucleus.  In this notation, $T$ is the recoil energy of the nucleus,
$M$ is its mass, and $F$ is its nuclear form factor (FF) evaluated at
the squared momentum transfer of the process, $Q^2 = 2 M T$.  In our
calculations we have first used a Helm FF\footnote{The collaboration
  used a slightly different FF, taken from
  Ref.~\cite{Klein:1999qj}. However, we have checked that the results
  of the fit using their parametrization gives identical results to
  those obtained using the Helm FF.}
parametrization~\cite{Helm:1956zz}:
\begin{equation}
  \label{eq:Helm}
  F(Q^2) = 3 \frac{j_1(Q R_0)} {Q R_0} e^{-Q^2 s^2 / 2}
\end{equation}
where $s = 0.9$~fm~\cite{Lewin:1995rx} and $j_1(x)$ is the order-1
spherical Bessel function of the first kind. The value of $R_0$
relates to the value of $s$ and the neutron radius $R_n$ as
\begin{equation}
  \label{eq:Rn}
  \frac{R_0^2}{5} = \frac{ R_n^2 }{3} - s^2 \,.
\end{equation}
In the absence of an experimental measurement of the neutron radius in
CsI, we tune its value so that the prediction for the total number of
events at COHERENT matches the official one provided in
Refs.~\cite{Akimov:2018vzs, Akimov:2017ade} (173 events). However, by
doing so we obtain $R_n = 4.83$~fm, a value that is unphysical as it
approaches the proton radius~\cite{Fricke:1995zz} which all models
predict to be smaller.  Given that this is a phenomenological
parametrization, though, and seeing the large differences in the
prediction for the total number of events obtained with different
values of $R_n$, it is worth asking whether this is accurate enough
for CsI, and exploring the impact of the nuclear FF on the results of
the fit. Therefore, in Sec.~\ref{sec:results} we will also show the
results obtained using a state-of-the-art theoretical calculation for
the nuclear FF, taken from Refs.~\cite{menendez, Klos:2013rwa}
(calculated using the same methodology as in
Refs.~\cite{Hoferichter:2016nvd, Hoferichter:2018acd}).

In the SM, the weak charge of a nucleus only depends on the SM vector
couplings to protons ($g_p^V$) and neutrons ($g_n^V$) and is
independent of the neutrino flavor:
\begin{equation}
  \Qwq \equiv \big( Z g_p^V + N g_n^V \big)^2 \,,
\end{equation}
where $g_p^V = 1/2 - 2\sin^2\theta_w$ and $g_n^V = -1/2$, with
$\theta_w$ being the weak mixing angle.  For CsI, we obtain $\Qwq
\simeq 1352.5$ in the SM.
However, in presence of NC NSI, this effective charge gets
modified\footnote{In practice, unless the ratio of the new couplings
  to up and down quarks remains the same as in the SM, the FF of the
  nucleus would also be affected by the NP and should be recomputed
  including the NP terms. However, in the case of vector-vector
  interactions (as in the case of NSI) the modifications to the
  nuclear FF are expected to be subleading, and the factorization of
  the NP effects into the weak charge approximately holds.  We warmly
  thank Martin Hoferichter for pointing this out.}  by the new
operators introduced as~\cite{Barranco:2005yy}:
\begin{multline}
  \label{eq:Qalpha-nsi}
  \Qwq_\alpha(\vec\Eps) = \left[
    Z \big(g_p^V + 2\Eps_{\alpha\alpha}^u + \Eps_{\alpha \alpha}^d \big)
    + N \big( g_n^V + \Eps_{\alpha\alpha}^u + 2\Eps_{\alpha \alpha}^d \big)
    \right]^2
  \\
  + \sum_{\beta \neq \alpha} \left[
    Z \big( 2\Eps_{\alpha \beta}^u + \Eps_{\alpha \beta}^d)
    + N \big( \Eps_{\alpha\beta}^u + 2\Eps_{\alpha\beta}^d \big)
    \right]^2 ,
\end{multline}
and in general its value may now depend on the NSI parameters
$\vec\Eps \equiv \{ \Eps_{\alpha\beta}^f \}$ as well as the incident
neutrino flavor $\alpha$. Since the COHERENT experiment observes
interactions of both electron and muon neutrinos, its results are
sensitive to both $\Qwq_e$ and $\Qwq_\mu$.

As can be seen from Eq.~\eqref{eq:Qalpha-nsi}, the modification of NSI
to the CE$\nu$NS event rate comes in as a normalization
effect. Therefore, adding energy information to the analysis of the
data is not expected to have a significant effect on the results of
our fit to NSI.\footnote{In principle, a subleading effect can be
  observed for experiments with large statistics, due to the different
  maximum recoil energies expected for the prompt and delayed neutrino
  components of the beam~\cite{Baxter:2019mcx}. However, we find that
  the COHERENT experiment is insensitive to this effect with the
  current exposure.} Conversely, the addition of timing information is
crucial as it translates into a partial discrimination between
neutrino flavors, thanks to the distinct composition of the prompt
($\nu_\mu$) and delayed ($\bar\nu_\mu$ and $\nu_e$) neutrino
flux. This translates into an enhanced sensitivity to NSI, since the
fit will now be sensitive to a change in normalization affecting
neutrino flavors differently.

\section{Implementation of the COHERENT experiment}
\label{sec:coh2}

In our previous work~\cite{Coloma:2017ncl} we performed a fit to
COHERENT data using the available information at that time, which
included only the total event rates observed.  Last year the
collaboration released publicly both the energy and timing information
of the events~\cite{Akimov:2018vzs}. In this section we describe the
procedure used to implement in our fit the information provided in
such data release.

\subsection{Computation of the signal}

The differential event distribution at COHERENT, as a function of the
nuclear recoil energy $T$, reads
\begin{equation}
  \label{eq:dNdT}
  \frac{dN}{dT} = N_\text{pot} f_{\nu/p} N_\text{nuclei} \sum_\alpha
  \int_{E_\nu^\text{min}}^{m_\mu/2} \frac{d\sigma_\alpha}{dT}
  \frac{d\phi_{\nu_\alpha}}{dE_\nu} dE_\nu \,,
\end{equation}
where $N_\text{nuclei}$ is the total number of nuclei in the detector,
$N_\text{pot} = 1.76 \cdot 10^{23}$ is the total number of protons on
target considered, and $f_{\nu/p} = 0.08$ is the neutrino yield per
proton.  In Eq.~\eqref{eq:dNdT} the sum runs over all neutrino flux
components ($\nu_e$, $\nu_\mu$, $\bar\nu_\mu$), and the upper limit of
the integral is given by the end-point of the spectrum from pion DAR,
while the minimum neutrino energy that can lead to an event with a
nuclear recoil energy $T$ is given by
\begin{equation}
  E_\nu^\text{min} = \sqrt{\frac{M T}{2}} \,.
\end{equation}

At COHERENT, the observable that is actually measured is the number of
photo-electrons (PE) produced by an event with a certain nuclear
recoil. In fact, the nuclear recoil energy in CE$\nu$NS events is
typically dissipated through a combination of scintillation (that is,
ionization) and secondary nuclear recoils (that is, heat). While
secondary recoils are \emph{the} characteristic signal of a nuclear
recoil (as opposed to an electron recoil, which favors ionization
instead), their measurable signal is much smaller than that of
electron recoils. The ratio between the light yields from a nuclear
and an electron recoil of the same energy is referred to as the
Quenching Factor (QF).

Besides being a detector-dependent property, the QF may also depend
non-trivially on the recoil energy of the nucleus. In general, the
relation between PE and nuclear recoil $T$ can be expressed as:
\begin{equation}
  \label{eq:QF}
  \text{PE} = T \cdot \text{LY} \cdot \text{QF}(T) \,,
\end{equation}
where LY is the light yield of the detector (that is, the number of PE
produced by an electron recoil of one keV), and we have explicitly
noted that the QF may depend on the nuclear recoil energy. Therefore,
the expected number of events in a certain bin $i$ in PE space can be
computed as:
\begin{equation}
  \label{eq:Nevents}
  N_i = \int_{T(\text{PE}_i^\text{min})}^{T(\text{PE}_i^\text{max})}
  \frac{dN}{dT} dT \,,
\end{equation}
where the limits of the integral correspond to the values of $T$
obtained for the edges of the PE bin ($\text{PE}_i^\text{min}$,
$\text{PE}_i^\text{max}$) from Eq.~\eqref{eq:QF}.

In their analysis, the COHERENT collaboration adopted an
energy-independent QF throughout the whole energy range considered in
the analysis, between 5 and 30~keV~\cite{Akimov:2017ade}. Also, given
the tension observed between the different calibration measurements
available at the time, they assigned large error bars to the assumed
central value $\overline{\text{QF}} = 8.78\%$. Taking a central value
for the light yield $\overline{\text{LY}} = 13.348$~PE per keV of
electron recoil~\cite{Akimov:2018vzs}, this means that approximately
1.17~PE are expected per keV of nuclear recoil energy. Very recently,
however, the authors of Ref.~\cite{Collar:2019ihs} have re-analyzed
past calibration data used to derive this result. They concluded that
the tension between previous measurements was partially due to an
unexpected saturation of the photo-multipliers used in the calibration
and, after correcting for this effect, a much better agreement was
found between the different data sets.  This allows for a significant
reduction of the error bars associated to the QF, as well as for the
implementation of an energy-dependent QF.  On the other hand, the
COHERENT collaboration has not confirmed the claims of the authors of
Ref.~\cite{Collar:2019ihs}. After repeating the calibration
measurements, their new data still shows a good
agreement~\cite{PhilBarbeau} with the original measurements performed
by the Duke (TUNL) group~\cite{Akimov:2018vzs, Akimov:2017ade}.

\begin{figure}\centering
  \includegraphics[scale=0.48]{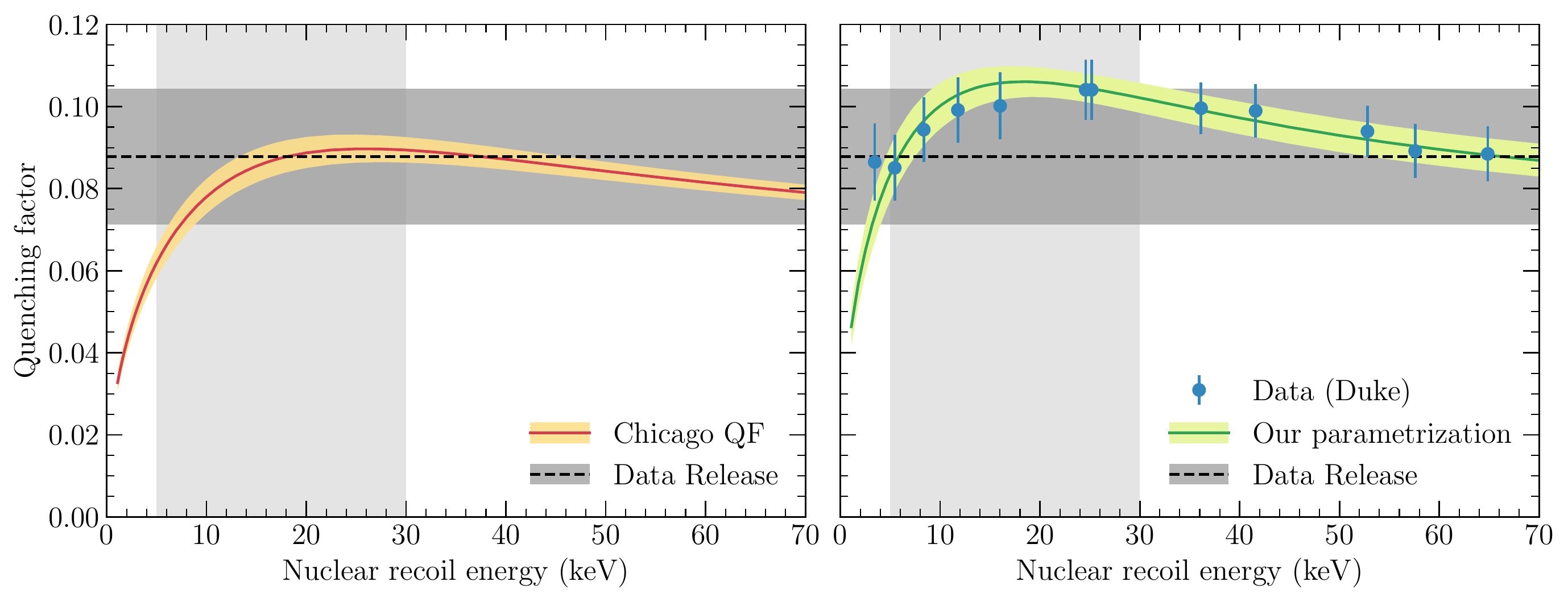}
  \caption{Quenching Factor parametrizations used in our analysis of
    the COHERENT data.  The left panel shows the curve provided in
    Ref.~\recite{Collar:2019ihs} (solid curve), while the right panel
    shows the corresponding result obtained for our fit to the
    calibration data of the Duke (TUNL) group~\recite{Akimov:2017ade},
    provided as part of the data release~\recite{Akimov:2018vzs}. In
    both panels, the constant QF used in Ref.~\recite{Akimov:2017ade}
    is also shown for comparison. The three parametrizations are shown
    with a shaded band to indicate the values allowed at the $1\sigma$
    CL in each case. For illustration, the vertical shaded area
    indicates approximately the range of nuclear recoil energies that
    enters the signal region used in the fit (the exact range varies
    with the nuisance parameters, and the exact QF parametrization
    used).}
  \label{fig:QF}
\end{figure}

Given that this issue has not been settled yet, we will study and
quantify the effect of these new measurements in the results of the
fit in Sec.~\ref{sec:results}. We will present our results obtained
for three different QF parametrizations: the original (constant)
parametrization used in the data release~\cite{Akimov:2018vzs}; the
best-fit obtained by the authors of Ref.~\cite{Collar:2019ihs}; and
the results from our fit to the calibration measurements performed by
the TUNL group~\cite{Akimov:2017ade, Akimov:2018vzs}. In order to fit
the data of the Duke group, we use the phenomenological
parametrization proposed in~\cite{Collar:2019ihs}, which is based on a
modification of the semi-empirical approach by
Birks~\cite{Birks:1951boa} and depends only on two parameters, $E_0$
and $\textit{kB}$.\footnote{In brief, the fitted functional form of
  the QF as a function of the nuclear recoil energy $T$ is
  $\text{QF}(T) = [1-\exp(-T/E_0)] / [\textit{kB}\, dE/dR(T)]$ where
  $dE/dR(T)$ is the energy loss per unit length of the ions. For
  simplicity, we take it to be the average between that of Cs and I,
  obtained from SRIM-2013~\cite{SRIM}. We have also verified that
  using a simple polynomial parametrization for $QF(T)$ leads to very
  similar results.}  Following this approach, we obtain a best-fit to
the Duke group data for $E_0 = 9.54 \pm 0.84$ and $\textit{kB} = 3.32
\pm 0.10$, with a correlation $\rho_{\textit{kB},E_0} = -0.69$. The
three QF parametrizations used in our calculations are shown in
Fig.~\ref{fig:QF}, as a function of the recoil energy of the nucleus.

Once the expected event distribution in PE has been computed following
Eq.~\eqref{eq:Nevents}, the expected number of events in each bin has
to be smeared according to a Poisson distribution, to account for the
probability that a given event yields a different number of PE than
the average. On top of that, signal acceptance efficiencies are
applied to each bin:
\begin{equation}
  \label{eq:acceptance}
  \eta(\text{PE}) = \frac{\eta_0}{1 + e^{-k (\text{PE} - \text{PE}_0)}} \,
  \Theta(\text{PE} - 5) \,,
\end{equation}
where the function $\Theta$ is defined as:
\begin{equation}
  \label{eq:Theta}
  \Theta(\text{PE} - 5) =
  \begin{cases}
    0 & \text{if~} \text{PE} < 5 \,,
    \\
    0.5 & \text{if~} 5 < \text{PE} < 6 \,,
    \\
    1 & \text{if~} 6 < \text{PE} \,.
  \end{cases}
\end{equation}
Following Ref.~\cite{Akimov:2018vzs}, the central values of the signal
acceptance parameters are set to $\bar\eta_0 = 0.6655$, $\bar{k} =
0.4942$, $\overline{\text{PE}}_0 = 10.8507$.

Finally, once the predicted energy spectrum has been computed, one
should consider the arrival times expected for the different
contributions to the signal. This is implemented using the
distributions provided by the COHERENT collaboration in the data
release~\cite{Akimov:2018vzs}, which are normalized to one.

\begin{figure}\centering
  \includegraphics[width=0.47\textwidth]{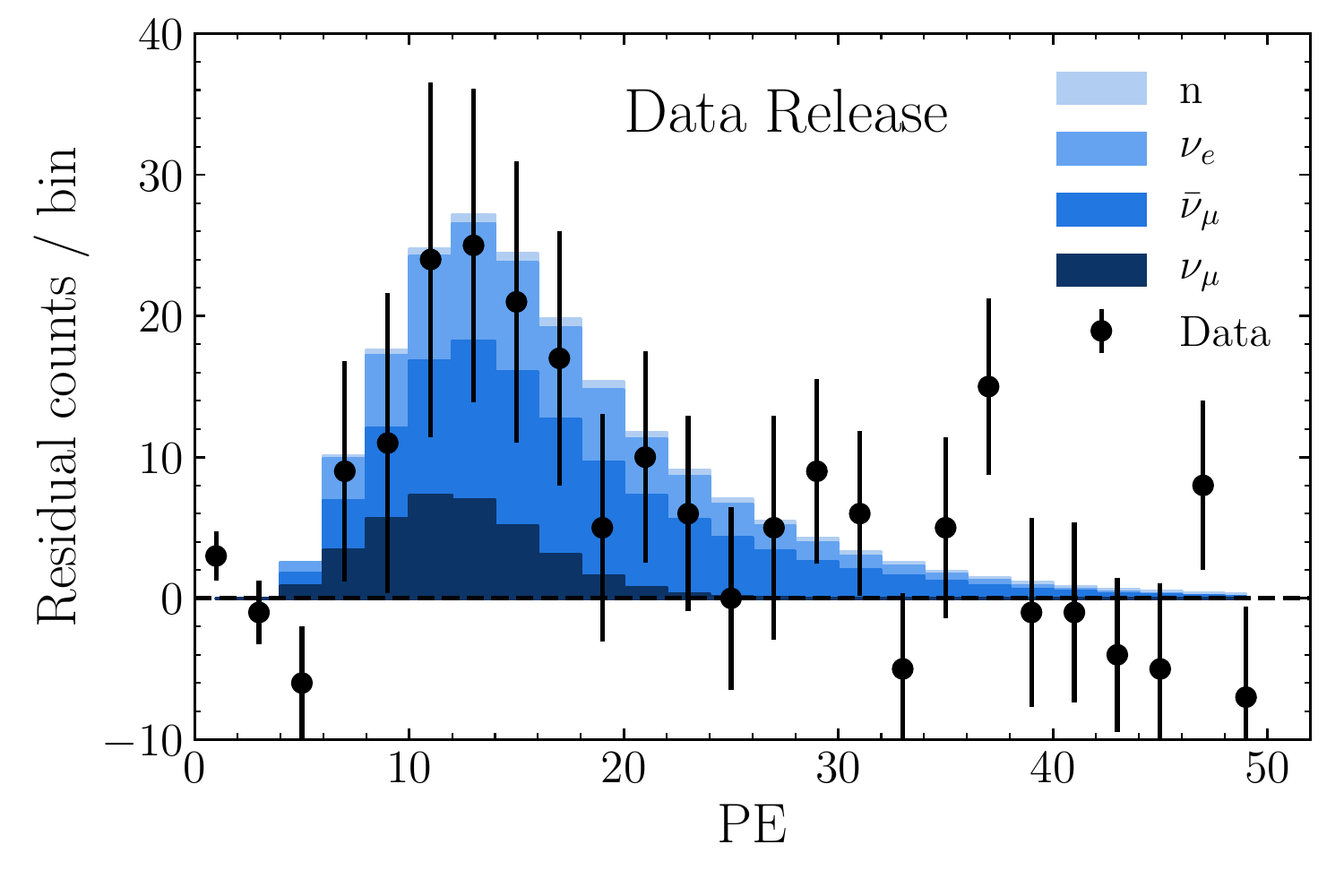}
  \includegraphics[width=0.47\textwidth]{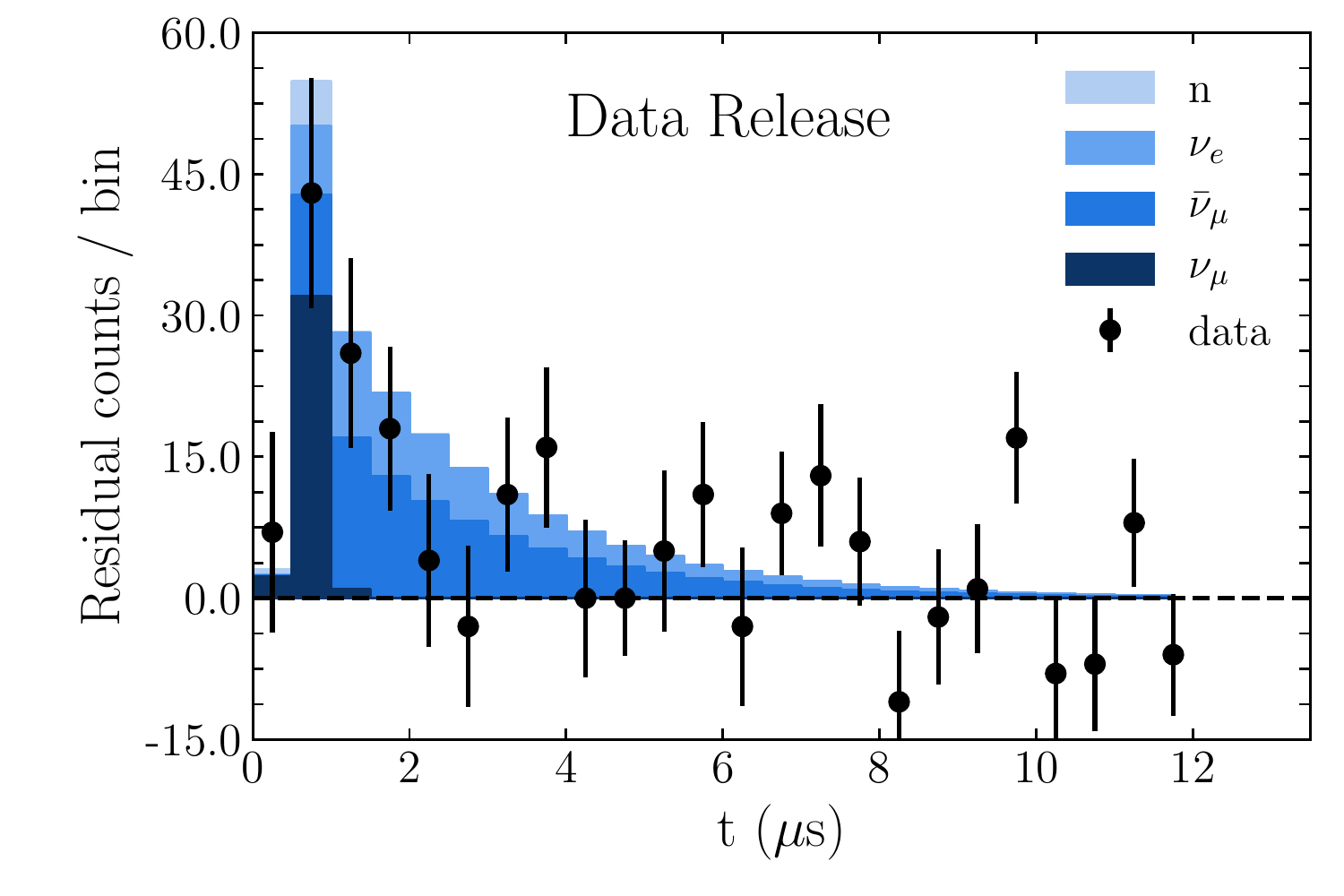}
  \includegraphics[width=0.47\textwidth]{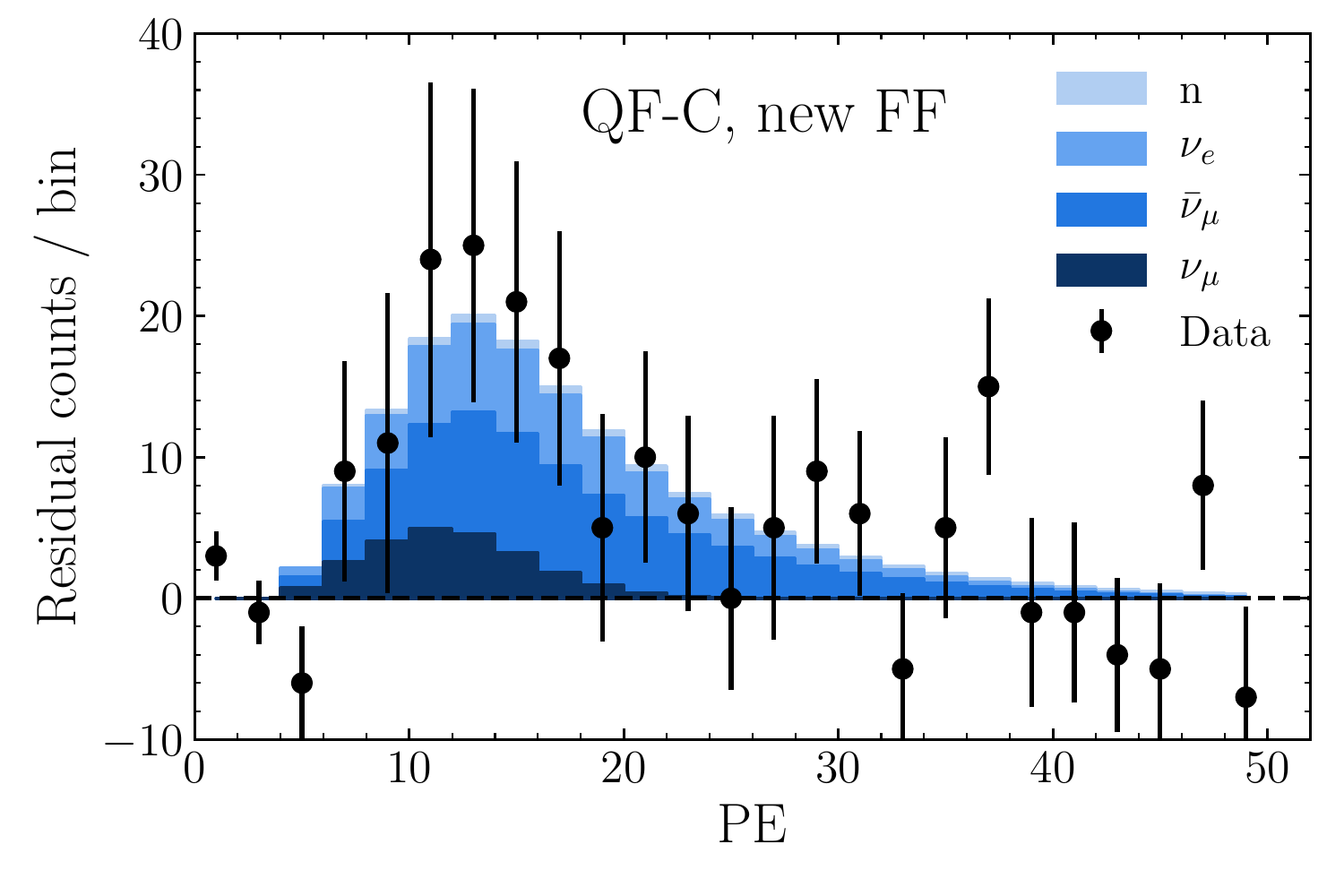}
  \includegraphics[width=0.47\textwidth]{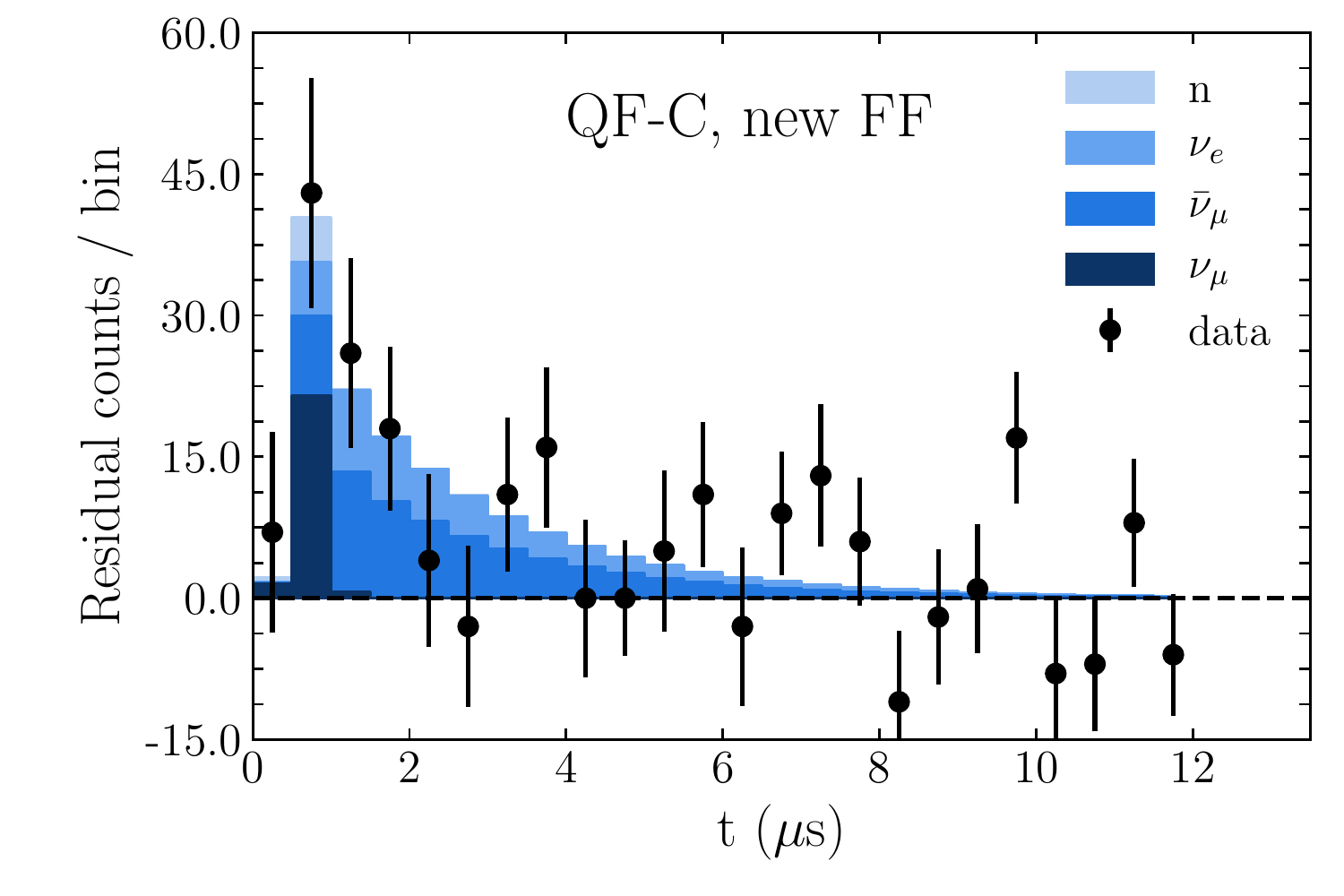}
  \includegraphics[width=0.47\textwidth]{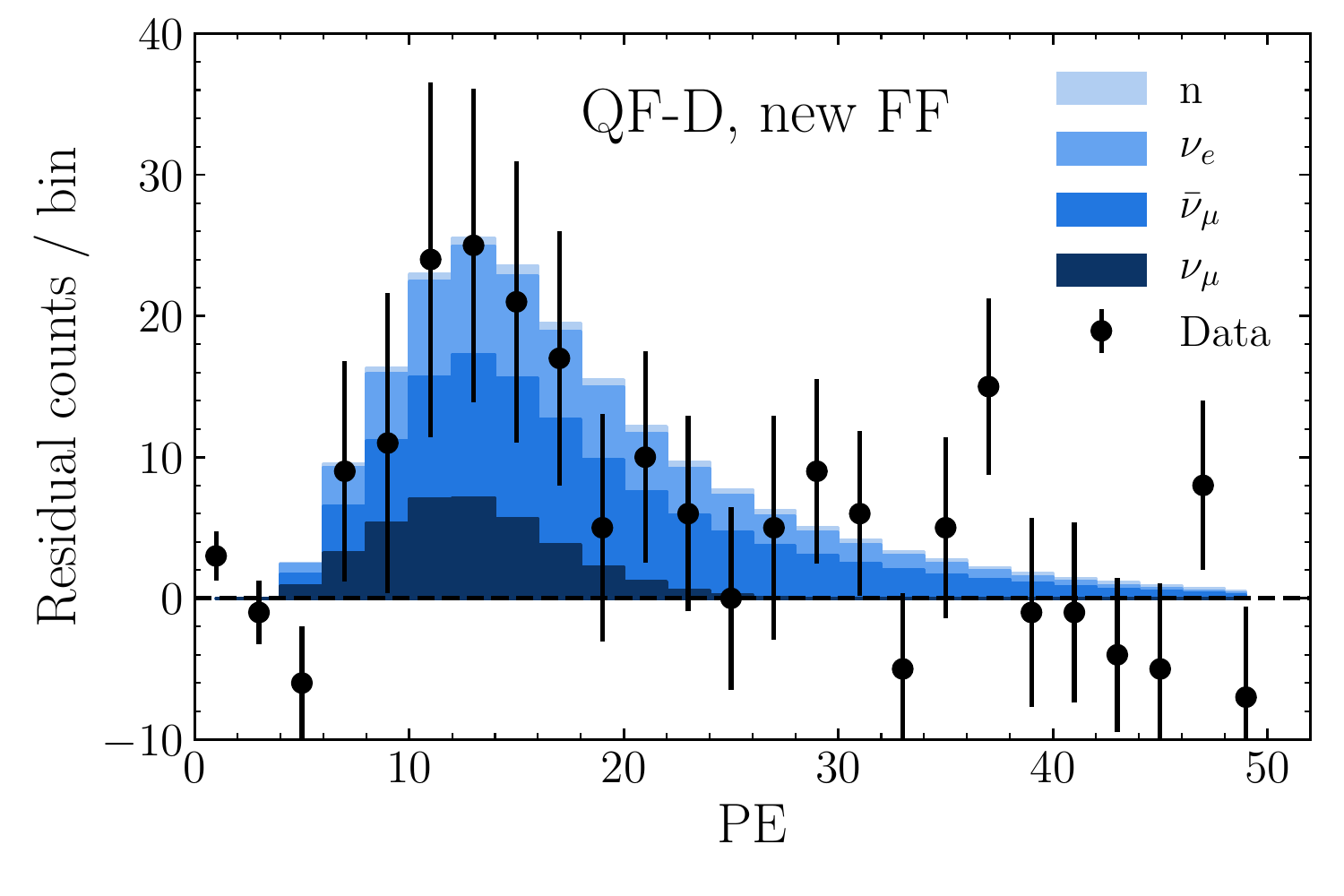}
  \includegraphics[width=0.47\textwidth]{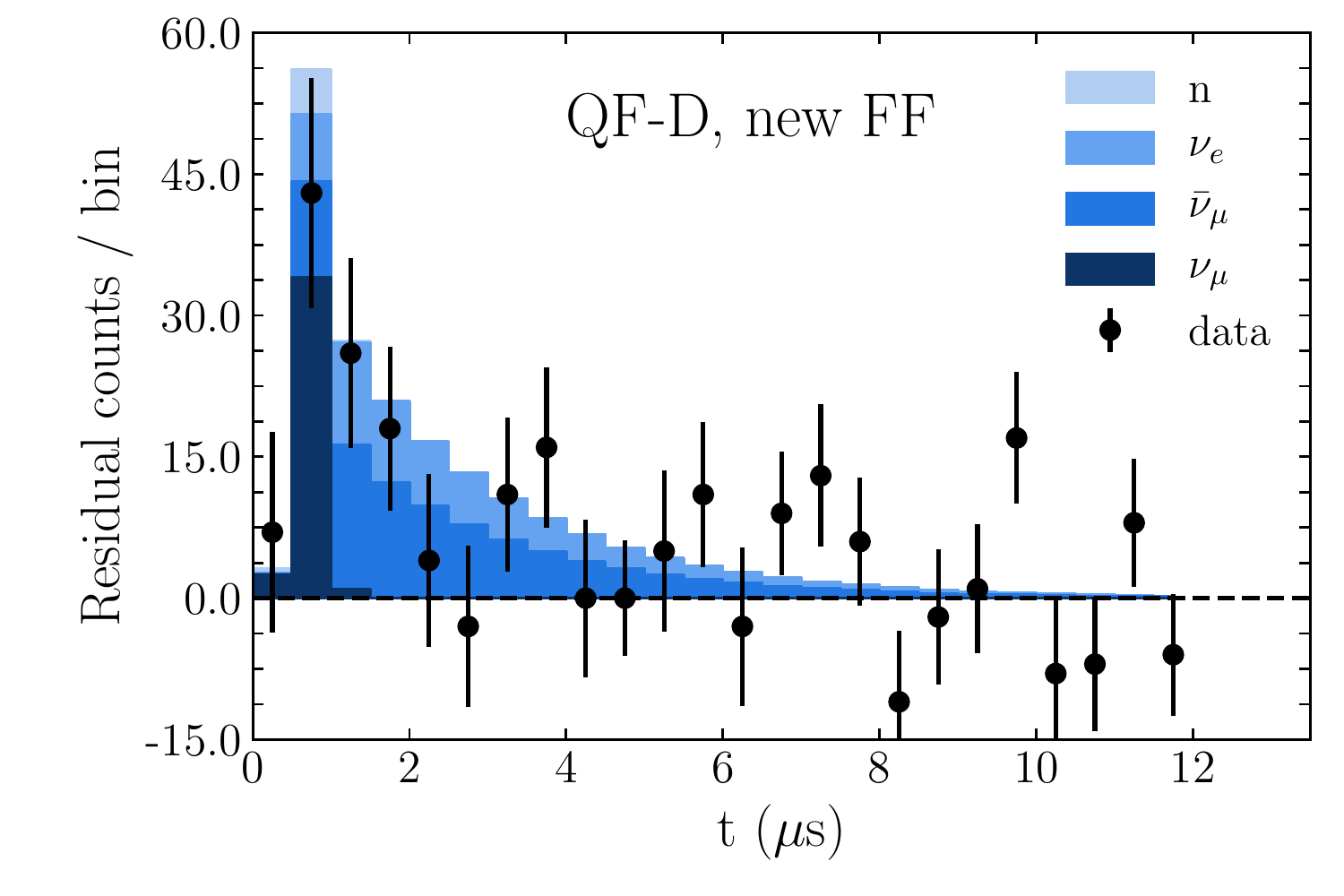}
  \caption{Residual events per bin obtained after subtracting C and AC
    data for the beam-ON sample, after being projected onto the PE
    (left panels) and time (right panels) axes and using the same cuts
    in PE an time as those applied to Fig.~3 in
    Ref.~\recite{Akimov:2017ade}. The observed data points are
    indicated with statistical error bars, as in
    Ref.~\recite{Akimov:2017ade}. In the upper panels the shaded
    histograms show the predicted event rates in the SM using the QF
    and nuclear FF from Ref.~\recite{Akimov:2018vzs}. In the middle
    panels they correspond to the predictions with the QF from the
    Chicago group (QF-C) in Ref.~\recite{Collar:2019ihs} and the
    nuclear FF from Ref.~\recite{menendez,Klos:2013rwa}.  The lower
    panels have been obtained with the same FF, but changing the QF to
    match the Duke (TUNL) measurements in Ref.~\recite{Akimov:2017ade}
    (QF-D).  In all panels the prompt neutron background prediction is
    also shown for completeness. All the event histograms shown in
    this figure correspond to the SM prediction.}
  \label{fig:histo-res}
\end{figure}

This final prediction can be compared with the published data.  This
is provided in two different time windows for each trigger in the data
acquisition system (that is, for each proton pulse). On the one hand,
the region where signals and beam-induced backgrounds associated with
the SNS beam are expected is referred to as the coincidence (C)
region, which can therefore be considered a ``signal'' region.
Conversely, the region where no contribution from the SNS beam is
expected is referred to as the anti-coincidence (AC) region and could
be considered a ``background'' region. While the collaboration
provides data separately for the beam-ON and beam-OFF data taking
periods, in this work we only use the beam-ON samples. The total
exposure considered in this work corresponds to 308.1 live-days of
neutrino production, which correspond to 7.48 GW-hr ($\sim 1.76\times
10^{23}$ protons on target). The residual event counts for this
period, \textit{i.e.}, the C data with the AC data subtracted, are
shown in Fig.~\ref{fig:histo-res}, projected onto the time and PE
axes, for different choices of the QF and FF as indicated by the
labels.

As can be seen from the comparison between the upper and lower panels,
the change in QF between a constant approximation (Data Release) and
the energy-dependent result obtained by the TUNL group (QF-D) does not
affect significantly the predicted event distributions. This will lead
to a minor change in the results of the numerical fit to the data in
Sec.~\ref{sec:results}. A larger difference is observed with respect
to the predictions using the QF by the Chicago group (QF-C, middle
panels): in this case, the very different central values at $T \sim
10$~keV (corresponding to $\text{PE}\sim 10$) lead to a reduced number
of events, which will have a larger impact on the results.

\subsection{Computation of the background}
\label{sec:bg}

The COHERENT measurement is affected by three main background sources:
(i) the steady-state background, coming from either cosmic rays or
their by-products entering the detector; (ii) prompt neutrons produced
in the target station and exiting it, and (iii) neutrino-induced
neutrons (NINs) that originate in the shielding surrounding the
detector. While the latter is irreducible, it has been shown to be
negligible at the COHERENT experiment and is therefore ignored here.

The procedure used to compute the expected number of background events
for the steady-state and the prompt neutron components follows the
prescription given in Ref.~\cite{Akimov:2018vzs}. For both
backgrounds, it is assumed that the temporal and energy dependence on
the number of events can be factorized as
\begin{equation}
  \label{eq:bg}
  N_\text{bg}(t, \text{PE}) = f(t) \cdot g(\text{PE}) \,,
\end{equation}
where $f$ contains the temporal dependence of the signal and $g$ its
energy dependence.

For the prompt neutron background, the collaboration provides both its
expected energy distribution before acceptance efficiencies are
applied (that is, $g(\text{PE})$), and the total expected counts as a
function of time (that is, $f(t)$). The expected 2D distribution can
be obtained simply by multiplying the two distributions. After the
number of events in each bin has been computed, the same acceptance
efficiency as for the signal, Eq.~\eqref{eq:acceptance}, is applied to
determine the expected number of events in each bin.

The steady-state is the most significant background source to this
analysis, and it has the largest impact on the fit. In this case, the
functions $g(\text{PE})$ and $f(t)$ are not provided in
Ref.~\cite{Akimov:2018vzs} but inferred from the data, which is
provided per bin in energy and time. In particular, the projected data
onto the PE axis is then used directly as $g(\text{PE})$, while $f(t)$
is assumed to follow an exponential:
\begin{equation}
  \label{eq:expo-fit}
  f_\text{ss}(t) = a_\text{ss} e^{-b_\text{ss} t} \,.
\end{equation}
By taking the AC data and projecting it into the time axis, a best-fit
to the steady-state background is obtained for $a_\text{ss} = 58.5$
and $b_\text{ss} = 0.062$. The value of $f(t)$ is then normalized so
that its integral over the whole range in time is equal to one. Since
in this case the expected background events are inferred from a
measurement, the signal acceptance has already been included into the
calculation and there is no need to apply it here.

\begin{figure}\centering
  \includegraphics[width=0.65\textwidth]{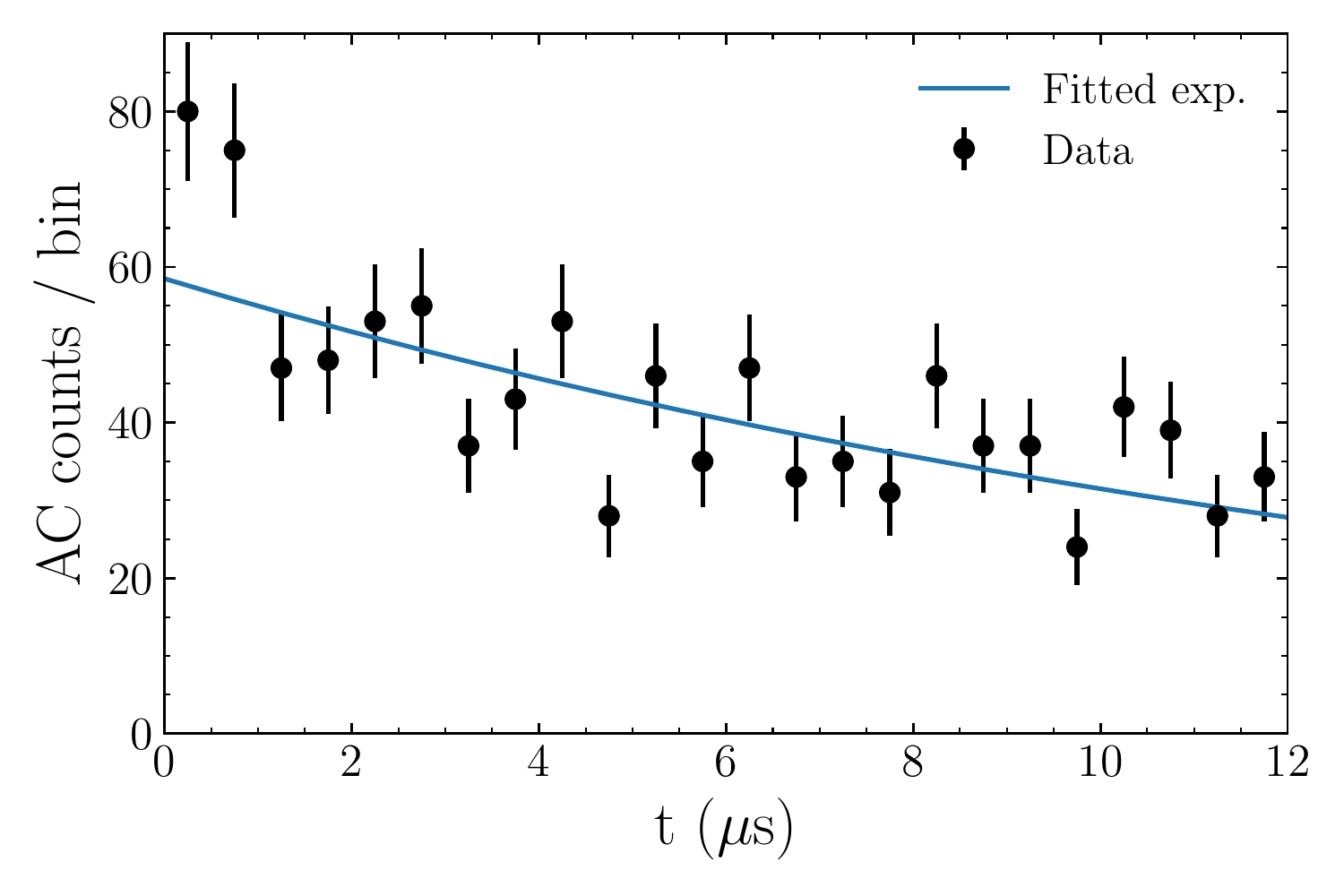}
  \caption{Total AC counts per bin, compared to the results of the
    exponential fit employed in Ref.~\recite{Akimov:2018vzs} and
    described in Sec.~\ref{sec:bg} used to model the steady-state
    background. No cuts on the observed number of PE have been applied
    to this figure.}
  \label{fig:bckg}
\end{figure}

The procedure outlined above for the steady-state component is meant
to eliminate biases in the fit due to the limited statistics of the
data sample used. However, by treating the background in this way the
analysis is rather sensitive to a mis-modeling of its temporal
component. In particular, if we plot the separate C and AC event
distribution as a function of time, instead of looking at their
difference, it is easy to see that there is an excess in the first two
bins in the data, which cannot be accommodated by the simplified
exponential fit. This is shown in Fig.~\ref{fig:bckg}, where we show
the total AC counts (which should include only the steady-state
background as measured by the detector) together with the exponential
that gives a best fit to the data. As clearly seen, the first two bins
are not well fitted by a simple exponential model.

\begin{figure}\centering
  \includegraphics[width=0.48\textwidth]{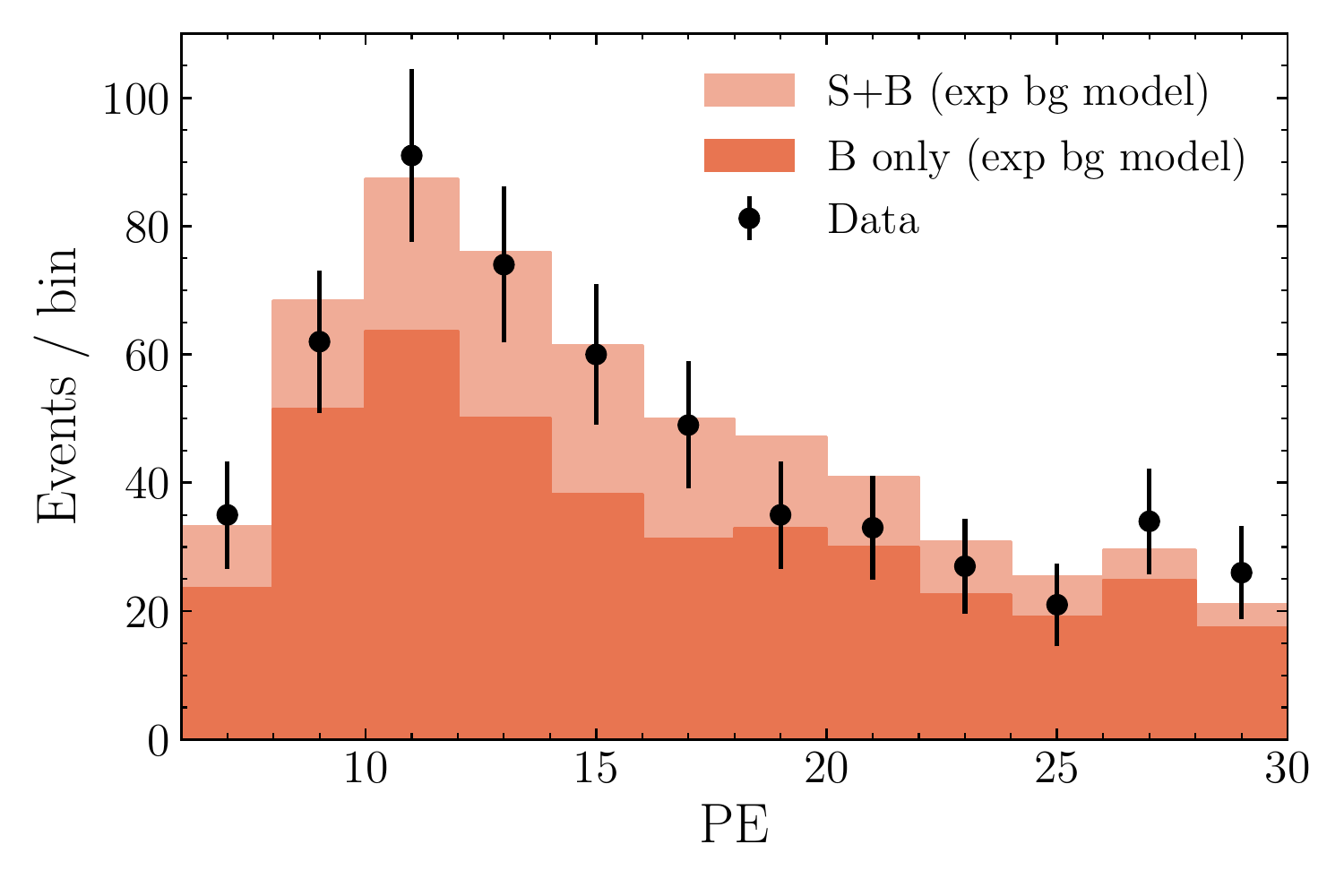}
  \includegraphics[width=0.48\textwidth]{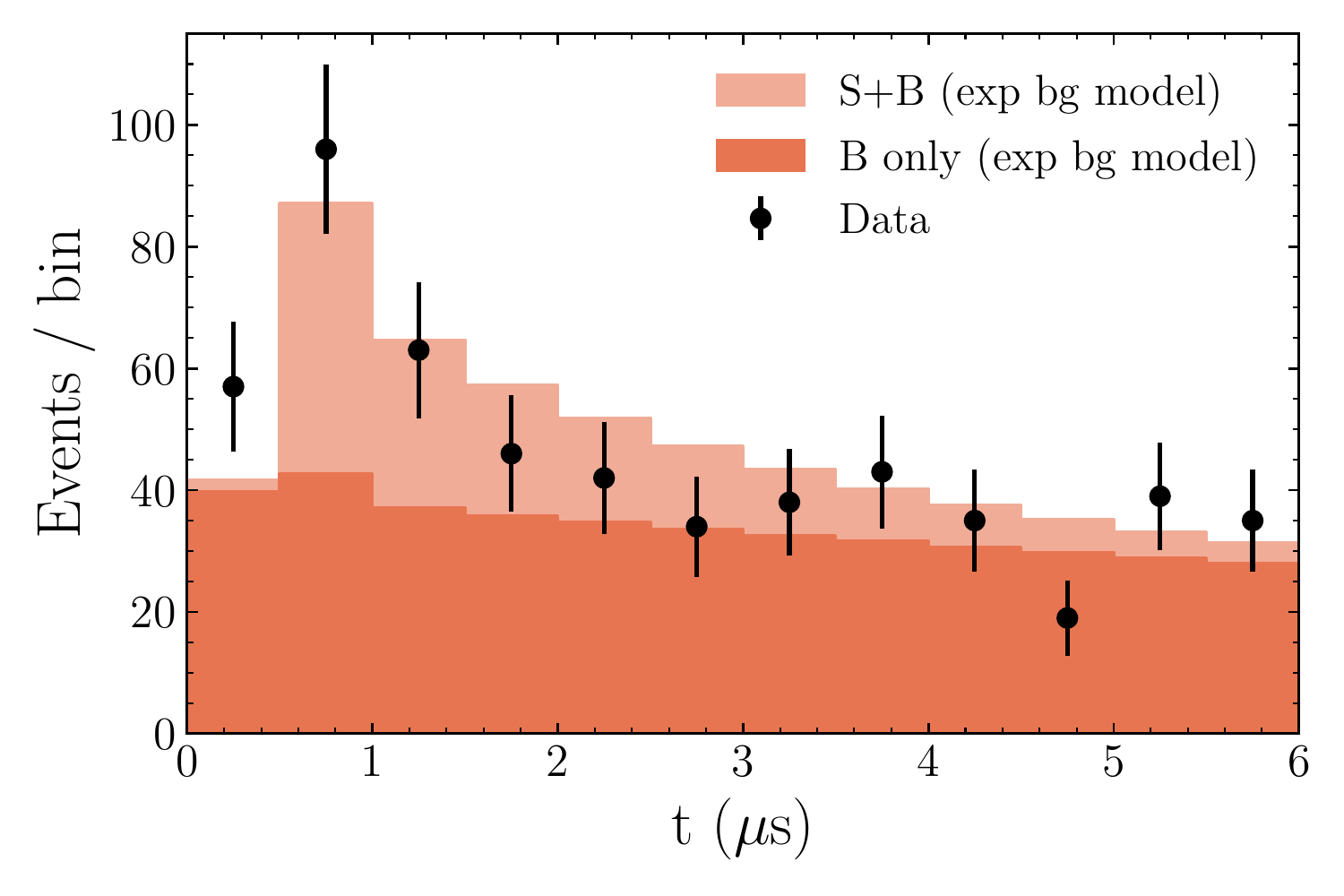}
  \includegraphics[width=0.48\textwidth]{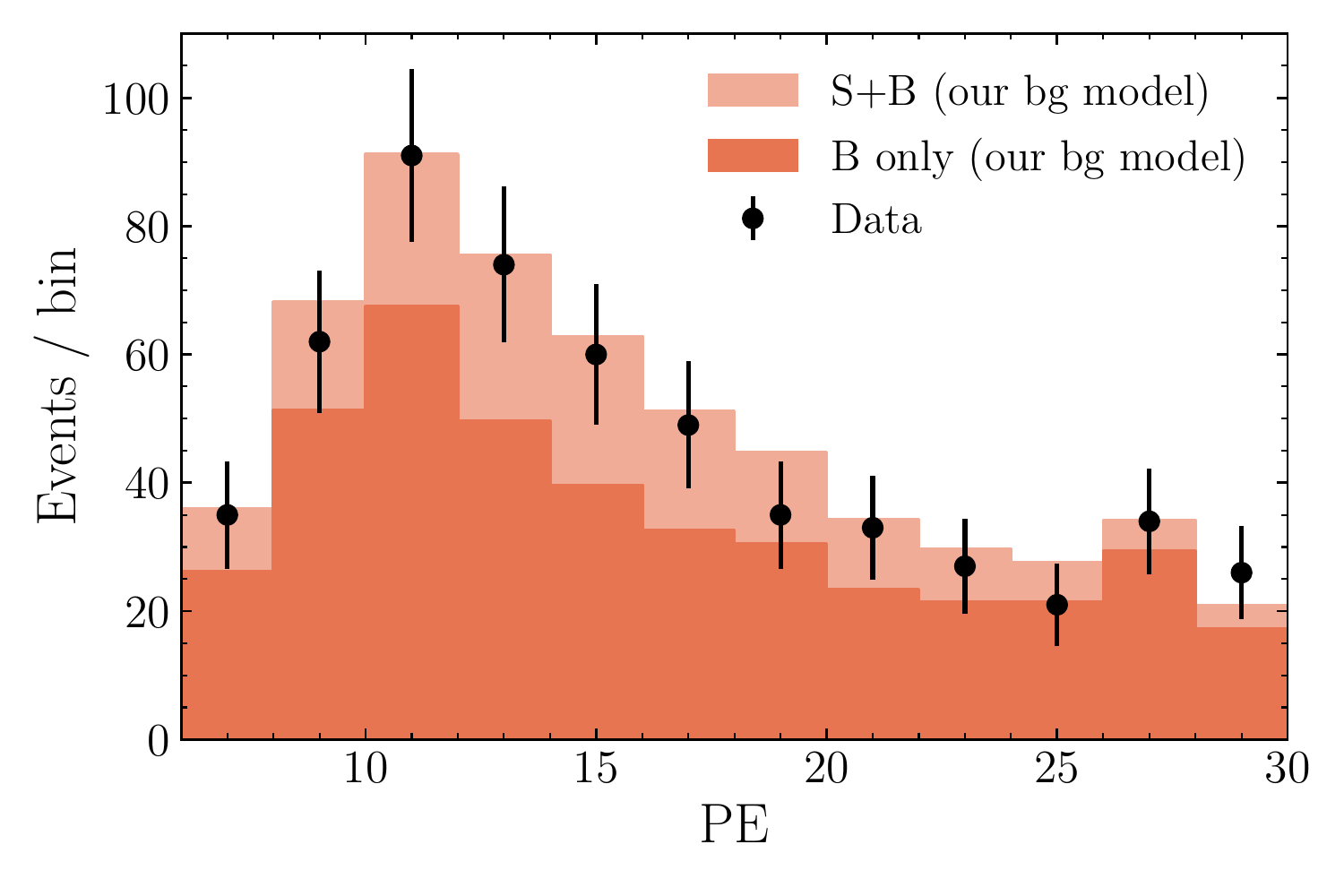}
  \includegraphics[width=0.48\textwidth]{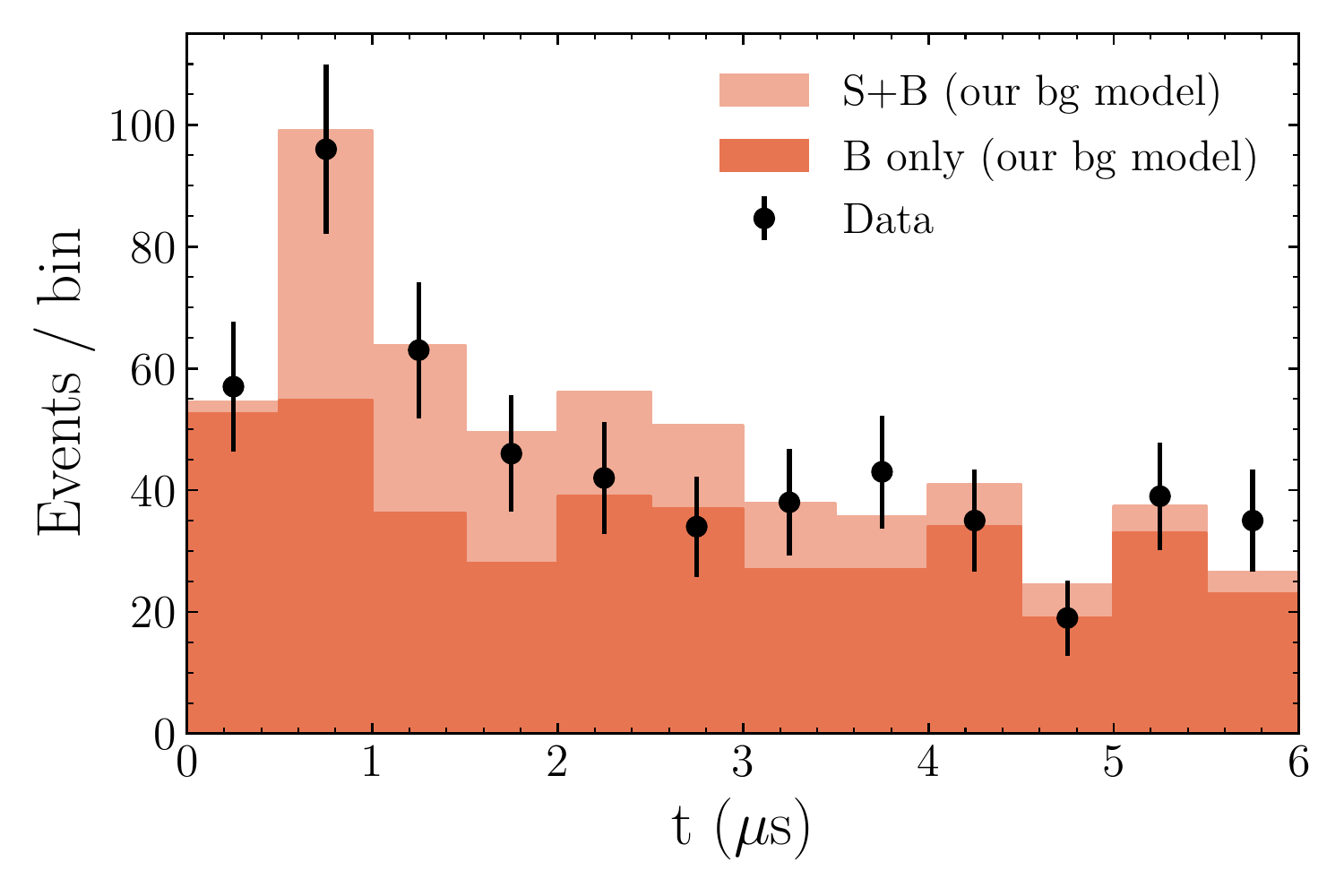}
  \caption{Total events per bin in the beam-ON C sample, after being
    projected onto the time (left) and PE (right) and imposing the
    cuts $5 < \text{PE} \leq 30$ and $t < 6~\mu\text{s}$. The observed
    data points are indicated with statistical error bars as in
    Fig.~\ref{fig:histo-res}.  The dark histograms show the expected
    background events for the steady-state contribution only.  The
    upper panels have been obtained assuming that the time dependence
    of the background follows an exponential model, as in
    Ref.~\recite{Akimov:2018vzs}, while the lower panels have been
    obtained using our model (which follows the time dependence of the
    AC events, see text for details). The light histograms show the
    predicted total number of events, after adding all signal and
    background contributions. To ease the comparison among different
    panels, in this figure the signal has been computed in all cases
    using the same FF and QF as in the data
    release~\recite{Akimov:2018vzs}. All histograms shown correspond
    to the SM predicted event rates.}
  \label{fig:histo-total}
\end{figure}

Interestingly enough, both the C and AC samples seem to observe a
similar excess in the first two temporal bins, which suggests that
this contribution is not related to the neutrino signal but to some
mis-modeling of the background. A possible way to correct for this is
to directly use the measured time dependence of the AC sample as a
direct prediction for the expected behavior of the steady-state
background in the C sample. Doing this on a bin-per-bin basis would
not provide a good predictor for the expected number of events in each
bin, due to the limited statistics. However, the projected data onto
the time axis may still be used, as in the case of the exponential
fit, to get a prediction for the function $f(t)$. In other words, the
prediction for $f(t)$ may be obtained following the same procedure as
was done for $g(\text{PE})$, \textit{i.e.}, projecting the events onto
the corresponding axis.  Figure~\ref{fig:histo-total} shows the
observed total event counts for the beam-ON, C sample (which includes
both signal and background) projected onto PE (left) and time (right),
compared with the predictions using these two different background
models.

As can be seen from this figure, both background models are able to
reproduce the observed spectrum in PE relatively well, and give very
similar results. However, the event rates obtained with this second
method provide a better fit to the data when projected onto the time
axis and, as clearly observed from the figure, the effect is specially
noticeable in the first two bins. Therefore, in Sec.~\ref{sec:results}
we will show two sets of results: with and without using an
exponential model for the background.

\subsection{Systematic errors and implementation of the $\chi^2$}

Once the predicted event distributions for the signal and backgrounds
have been computed, a $\chi^2$ function is built as:
\begin{equation}
  \label{eq:chi2}
  \chi^2\big[P_{ij}(\vec\xi)\big]
  = \sum_{ij} 2 \left[ P_{ij}(\vec\xi) - O_{ij} +
    O_{ij} \ln\left( \frac{O_{ij}}{P_{ij}(\vec\xi)}\right) \right],
\end{equation}
where $O_{ij}$ stands for the observed number of events in PE bin $i$
and time bin $j$, while $P_{ij}$ stands for the total number of
predicted events in that bin, including the signal plus all background
contributions. Following Ref.~\cite{Akimov:2018vzs}, we consider only
the events with $5 < \text{PE} \leq 30$ and $t < 6~\mu\text{s}$ in the
analysis. The predicted number of events depends on the nuisance
parameters $\vec\xi \equiv \{ \xi_a \}$ included in the fit, which
account for the systematic uncertainties affecting the QF, signal
acceptance, neutrino production yield, and normalization of the
backgrounds. These are implemented replacing the original quantity as
$x \to (1 + \sigma_x \xi_x)\bar{x}$, where $\bar{x}$ denotes the
central value assumed for $x$ prior to the experiment and $\sigma_x$
denotes the relative uncertainty for nuisance parameter $\xi_x$
summarized in Tab.~\ref{tab:sys} for convenience.  More specifically:
\begin{multline}
  \label{eq:pij}
  P_{ij}(\vec\xi) = (1 + \sigma_\text{ss} \xi_\text{ss}) N_{ij}^\text{ss}
  \\
  + \eta(\text{PE}_i \,|\, \xi_{\eta_0}, \xi_k, \xi_{\text{PE}_0})
  \big[ (1 + \sigma_\text{n} \xi_\text{n}) N_{ij}^\text{n}
  + (1 + \sigma_\text{sig}\xi_\text{sig}) N_{ij}^\text{sig}(\xi_\text{QF})
  \big] ,
\end{multline}
where $N_{ij}^\text{ss}$, $N_{ij}^\text{n}$ and $N_{ij}^\text{sig}$
stand for the predicted number of events for the steady-state
background, the prompt neutron background and the signal. In
Eq.~\eqref{eq:pij} we have generically denoted as $\xi_\text{QF}$ the
set of nuisance parameters characterizing the uncertainty on the QF
employed. For the constant parametrization used in the data
release~\cite{Akimov:2018vzs} we introduce a unique nuisance parameter
with constant uncertainty.  For QF-C we introduce also a unique
nuisance parameter, but with an energy-dependent uncertainty inferred
from the uncertainty band in Fig.~1 of Ref.~\cite{Collar:2019ihs}
(also shown in the left panel of Fig.~\ref{fig:QF}), which varies from
6.5\% to 3.5\% in the range of recoil energies relevant for COHERENT.
For QF-D we introduce two nuisance parameters characterizing the
uncertainty on parameters $E_0$ and $\textit{kB}$ with their
corresponding correlation.\footnote{We have verified that, in
  practice, it is equivalent to using a single nuisance parameter with
  an energy-dependent uncertainty ranging between 8\% and 3\%
  (corresponding to the shaded band shown in the right panel in
  Fig.~\ref{fig:QF}).}

Altogether the likelihood for some physics model parameters
$\vec\Eps$, leading to a given set of predictions
$P_{ij}^{\vec\Eps}(\vec\xi)$ for the events in bin $ij$, is obtained
including the effects of the nuisance parameters as in
Eq.~\eqref{eq:pij} and minimizing over those within their assumed
uncertainty. This is ensured by adding a pull term to the $\chi^2$
function in Eq.~\eqref{eq:chi2} for each of the nuisance parameters
introduced:
\begin{equation}
  \label{eq:chi2min}
  \chi^2_\text{COH} (\vec\Eps)
  = \min\limits_{\vec\xi}
  \bigg\{ \chi^2\big[ P_{ij}^{\vec\Eps}(\vec\xi) \big]
  + \sum_{ab} \xi_a (\rho^{-1})_{ab} \xi_b \bigg\} \,,
\end{equation}
where $\rho$ is the correlation matrix, whose entries are $\rho_{ab} =
\delta_{ab}$ for all parameters except those entering our
parametrization of the QF of the Duke group (the corresponding
correlation coefficient can be found in Tab.~\ref{tab:sys}).

\begin{table}\centering
  \catcode`!=\active\def!{\hphantom{-}}
  \catcode`?=\active\def?{\hphantom{0}}
  \begin{tabular}{lc}
    \hline
    Parameter & Uncertainty (\%) \\
    \hline
    Steady-state norm. & $!?5.0?$ \\
    Prompt n norm. & $!25.0?$ \\
    Signal norm. & $!11.2?$ \\
    $\eta_0$ & $!?4.5?$ \\
    k & $!?4.7?$ \\
    $\text{PE}_0$ & $!?2.7?$ \\
    QF (data release) & $!18.9?$ \\
    QF (our fit, QF-C)& $!?6.5 - 3.5?$ \\
    $\textit{kB}$ (our fit, QF-D) &  $!?3.0?$    \\
    $E_0$ (our fit, QF-D) & $!?8.8?$   \\
    $\rho_{E_0,\textit{kB}}$ (our fit, QF-D) & $?{-0.69}$ \\
    \hline
  \end{tabular}
  \caption{Systematic uncertainties considered in the fit on
    acceptance efficiency parameters (Eq.~\eqref{eq:acceptance}),
    normalization of the signal and background contributions, and the
    QF. The steady-state normalization uncertainty includes the
    statistical error of the sample (AC data). The quoted
    uncertainties on the QF also includes the error on the light yield
    (0.14\%), which is however subdominant. For details on the QF
    parametrization, see Sec.~\ref{sec:coh2}.}
  \label{tab:sys}
\end{table}

\begin{figure}\centering
  \includegraphics[width=0.7\textwidth]{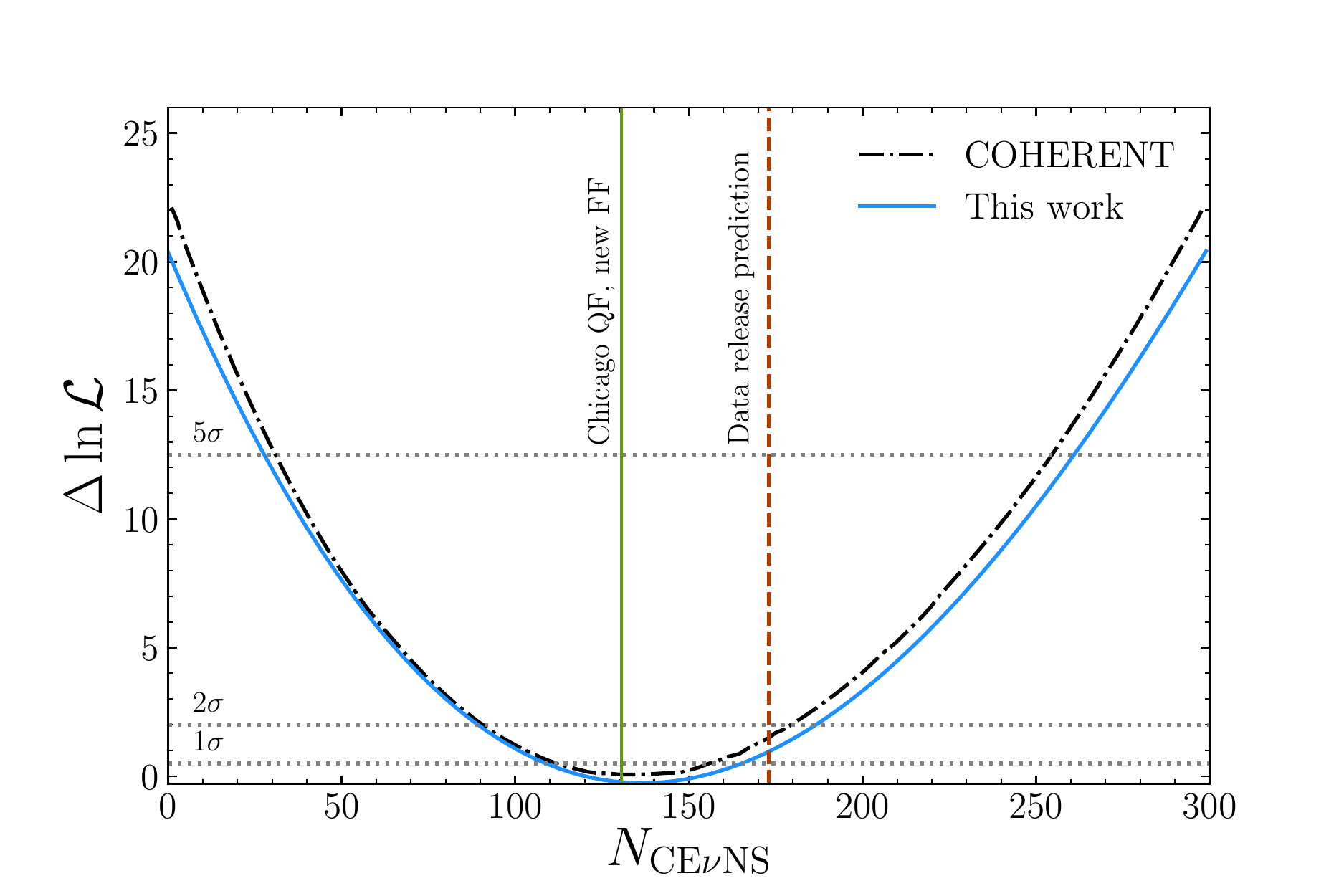}
  \caption{Comparison of our $\chi^2$ for the COHERENT timing and
    energy data as a function of the number of signal CE$\nu$NS events
    under the same assumptions on the background, systematics and
    expected time and energy dependence of signal, compared to that
    provided by COHERENT in figure S13 of
    Ref.~\recite{Akimov:2017ade}. For comparison, the vertical lines
    indicate the predicted event rates in the SM (with no systematic
    uncertainties), for different choices of QF and FF used: dashed
    red corresponds to the prediction provided in the data release of
    173 events~\recite{Akimov:2018vzs}, while solid green indicates
    our prediction using the QF from Ref.~\recite{Collar:2019ihs}
    (left panel in Fig.~\ref{fig:QF}) and nuclear FF
    of~\recite{menendez, Klos:2013rwa}.}
  \label{fig:compachi2}
\end{figure}

As validation of our $\chi^2$ construction we have performed a fit to
extract the total number of signal CE$\nu$NS events when using the
same assumptions on the background, systematics and energy and time
dependence of the signal as those employed by COHERENT in their data
release~\cite{Akimov:2018vzs}.  This can be directly compared with
their corresponding likelihood extracted from Figure S13 of
Ref.~\cite{Akimov:2017ade}. The result of this comparison is shown in
Fig.~\ref{fig:compachi2}.  Strictly speaking, the $\chi^2$ function
plotted in Fig.~\ref{fig:compachi2} depends on the assumed energy
dependence of the signal. Therefore it is expected to vary if, instead
of using the QF and FF quoted in the data release, we employed a
different QF parametrization and nuclear FF.  Quantitatively, within
the systematic uncertainties used in the construction of the shown
$\chi^2(N_\text{CE$\nu$NS}$), we find that changing the QF and FF has
a negligible effect on this curve.  Conversely, we find a stronger
dependence on the systematic uncertainties introduced, and therefore
Fig.~\ref{fig:compachi2} serves as validation of our implementation
for these.  For illustration, we also indicate the predicted event
rates in the SM predicted by the collaboration (173 events) as well as
our result obtained using the Chicago QF
parametrization~\cite{Collar:2019ihs}, as shown in the left panel of
Fig.~\ref{fig:QF} and the new nuclear FF from Refs.~\cite{menendez,
  Klos:2013rwa}. In both cases the vertical lines correspond to the
prediction without accounting for systematic uncertainties. The
predicted result using the Duke QF and the new FF from
Refs.~\cite{menendez, Klos:2013rwa} is very similar to the one
obtained by the collaboration (168 events) and is therefore not shown
here.

\section{Results}
\label{sec:results}

In this section we present our results. First, in
Sec.~\ref{sec:rescoh} we provide the results of the new fit using
COHERENT data alone, which we compare with those obtained in our
previous work~\cite{Coloma:2017egw}. We discuss in detail the
improvements coming from the inclusion of energy and timing
information in the fit, and investigate the impact of the different
choices of quenching factor, nuclear form factor and background
implementation.

We then proceed to combine the COHERENT data with the global analysis
of oscillation data in Sec.~\ref{sec:resglob}. We will show two main
sets of results, which quantify: (1) the quality of the global fit
once NSI are allowed, compared to the fit obtained under the SM
hypothesis; and (2) the status of the LMA-D solution after the
inclusion of COHERENT energy and timing data in the global fit. Both
sets of results are presented for a wide range of NSI models (that is,
for different values of $\eta$). Finally, we also provide the allowed
ranges obtained for the Wilson coefficients for three particular NSI
models, assuming that the new mediator couples predominantly to either
up/down quarks or to protons.

\subsection{Fit to COHERENT data}
\label{sec:rescoh}

\begin{figure}\centering
  \includegraphics[width=\textwidth]{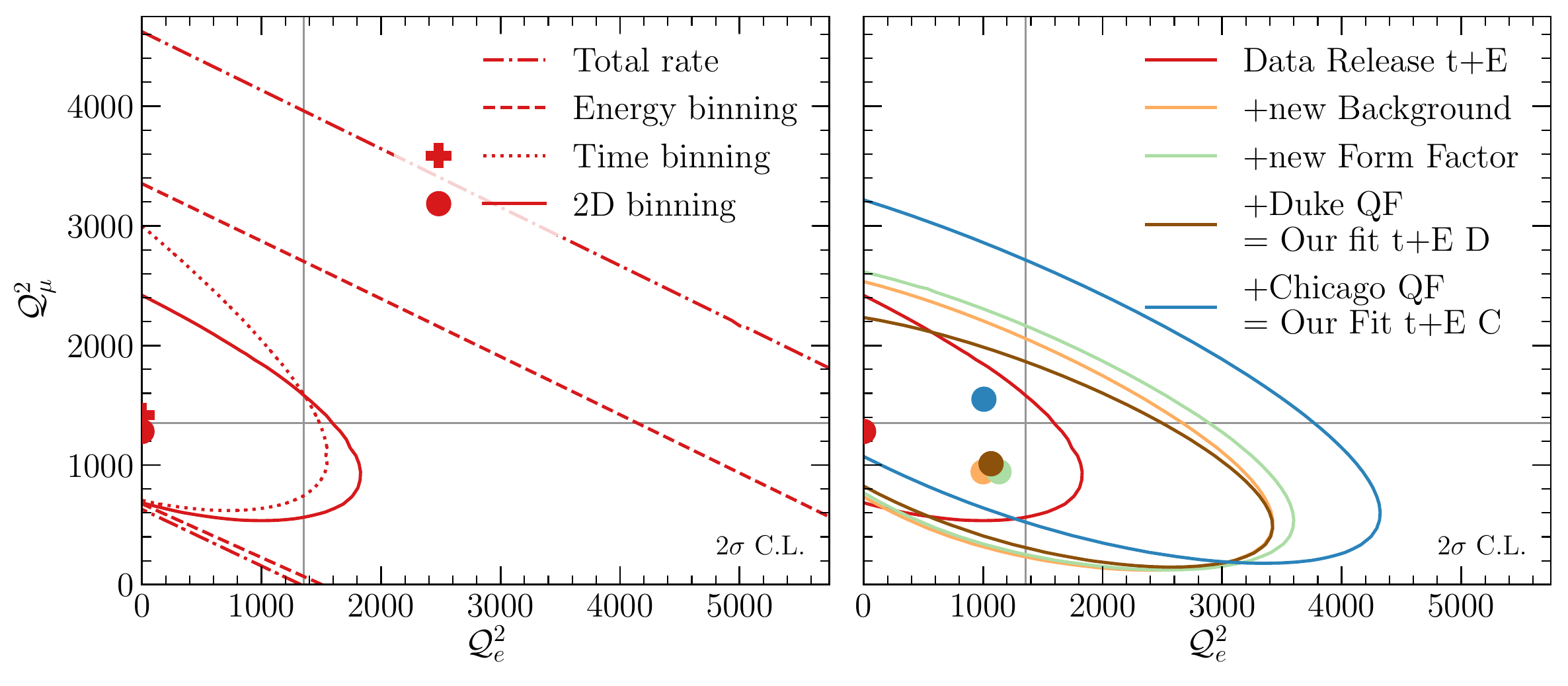}
  \caption{$2\sigma$ allowed regions for the flavor-dependent weak
    charges for a variety of fits to COHERENT data as labeled in the
    figure. In all cases shown in the left panel, the QF, nuclear FF
    and background assumptions are those employed in the data
    release~\recite{Akimov:2018vzs}.  On the right panel we show the
    dependence on the steady background modeling, nuclear FF, and
    QF. The vertical lines indicate the SM value $\Qwq_e = \Qwq_\mu =
    1353.5$. The colored dots and the red cross mark the position of
    the best-fit for the various cases.}
  \label{fig:qw}
\end{figure}

In order to study the dependence of the results on the different
assumptions, we have performed a set of fits to COHERENT data in terms
of two effective flavor-dependent weak charges $\Qwq_\alpha$ (assumed
to be energy-independent). Figure~\ref{fig:qw} shows the corresponding
allowed regions at $2\sigma$ from the fit to COHERENT data alone.  In
the left panel we illustrate the effect of including the energy and
timing information in the fit, by comparing the allowed values of the
weak charges obtained: (i) using only the total rate information
(dot-dashed); (ii) adding only the energy information (dashed); (iii)
using only the event timing information (dotted); and (iv) fitting the
data binned in both timing and energy (solid). In all cases shown in
the left panel, the QF, nuclear FF and background have been
implemented following closely the prescription given in the data
release~\cite{Akimov:2018vzs}. It is well-known that, when only the
total event rate information is considered, there is a degeneracy in
the determination of the flavor-dependent weak charges, since the
number of predicted events approximately behaves as
\begin{equation}
  \label{eq:Qbands}
  \Qwq_e f_{\nu_e} + \Qwq_\mu\, (f_{\nu_\mu} + f_{\bar\nu_\mu})
  \approx \frac{1}{3} \Qwq_e + \frac{2}{3} \Qwq_\mu
\end{equation}
where $f_\alpha$ indicates the fraction of expected SM events from
interactions of $\nu_\alpha$ in the final event sample and we have
assumed that one neutrino of each species is produced for each pion
DAR.  Under these assumptions, the allowed region in the $(\Qwq_e,
\Qwq_\mu)$ plane is a straight band with a negative slope,
$\arctan(-0.5) \approx -27^\circ$. This behavior is also observed from
our exact fit to the data, as shown by the dot-dashed lines in the
left panel in Fig.~\ref{fig:qw}.

As expected, the timing information is most relevant in breaking of
this degeneracy: since the prompt component of the beam contains only
$\nu_\mu$, the inclusion of time information allows for a partial
discrimination between $\Qwq_e$ and $\Qwq_\mu$.  Notice, however, that
in this case the best fit is obtained at the edge of the physically
allowed region, $\Qwq_e \simeq 0$ (in fact, it would probably take
place for a negative value, but this is not the case since we are
effectively imposing the restriction $\Qwq_\alpha > 0$ in the fit).
This is driven by the small excess for the event rates in the first
two time bins (with respect to the SM prediction) when using the
exponential fit model for the steady-state background, as described in
Sec.~\ref{sec:bg} (see Fig.~\ref{fig:histo-total}).  Such excess can
be accommodated thanks to the overall normalization uncertainty of the
signal, combined with a decrease of the $\nu_e$ contribution as
required to match the distribution observed for the delayed events.
Within the systematic uncertainties in the analysis, this results into
a higher rate at short times without a major distortion of the PE
spectrum.  We also observe that, including only the PE spectrum in the
fit, the degeneracy still remains but the width of the band in this
plane decreases.  For values of $\Qwq_\alpha$ in the non-overlapping
region, the fit using the event rate information alone is able to fit
the data, albeit at the price of very large nuisance parameters and,
in particular, of the QF-related uncertainties (which affect the shape
of the event distributions in PE space). Therefore, once the PE
information is added the allowed regions are consequently reduced.

The right panel in Fig.~\ref{fig:qw} shows the dependence of the
allowed region on the assumed background model, nuclear FF and QF
choice in the fit, for the 2D fit using both time and PE
information. As seen in the figure, if one uses the steady-state
background prediction without the exponential model for its temporal
dependence the region becomes considerably larger, and the BF moves
closer to the SM. This is expected because, with this background,
there is no excess of events in the first time bins with respect to
the SM prediction (see Fig.~\ref{fig:histo-total}). This also leads to
a better overall fit, with $\chi^2_\text{min} = 145.24$ (for $12\times
12=144$ data points) compared to $\chi^2_\text{min} = 150.8$ obtained
for the exponential model of the steady-state background.  Altogether
we observe that modifying the nuclear FF has a very small effect on
the current results, as can be seen from the comparison between the
orange and green lines in the figure.  Changing the QF does not have a
dominant impact either, once the exponential fit to the background has
been removed.  This can be seen from the comparison between the green,
brown and blue lines in the figure, which all provide similar results.
Overall, we find a slightly better agreement with the SM result for
the QF-C parametrization, albeit the effect is small.

\begin{figure}\centering
  \includegraphics[width=\textwidth]{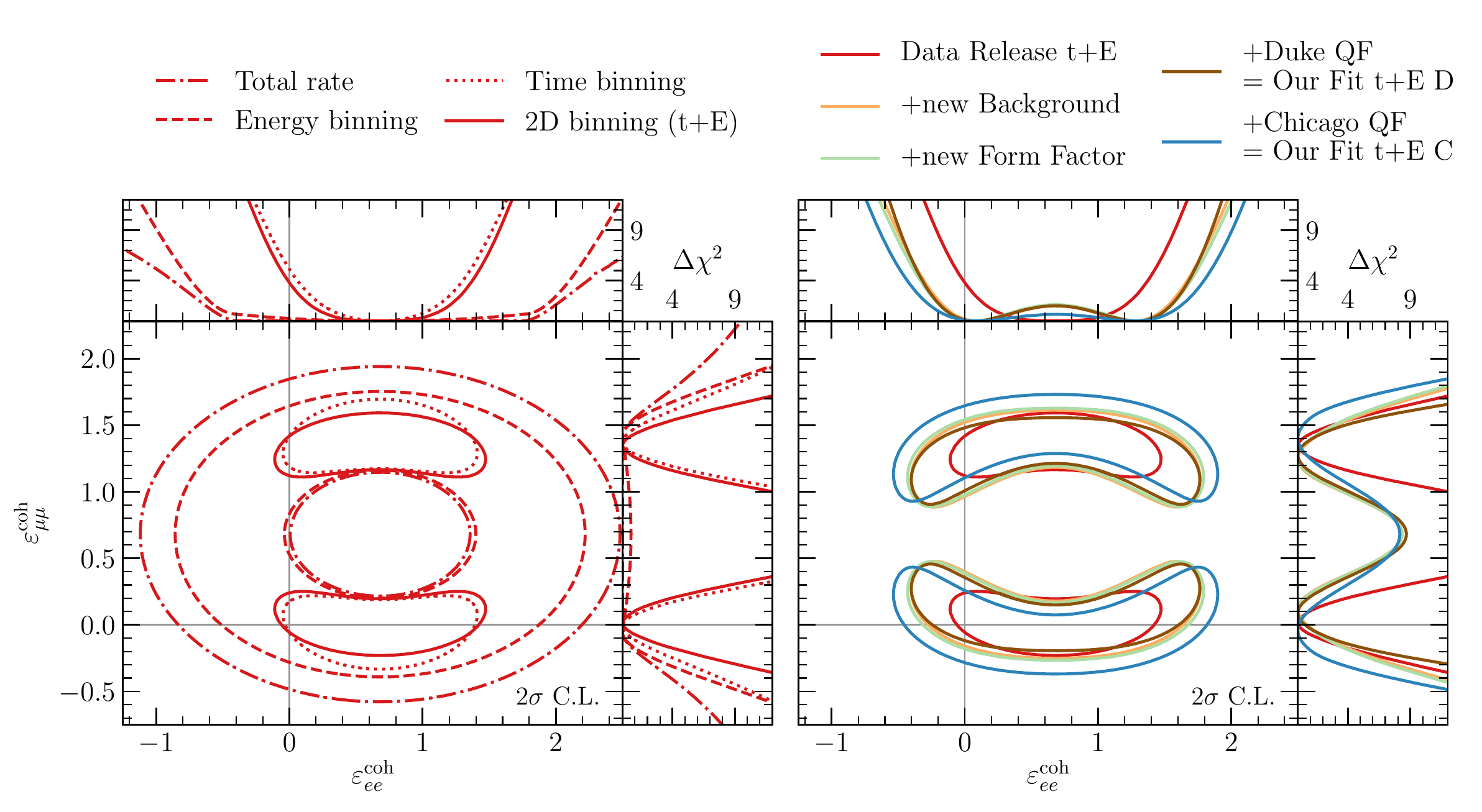}
  \caption{$2\sigma$ allowed regions for the flavor-diagonal NSI
    coefficients $\Eps_{\alpha\beta}^\text{coh}$ (assuming zero
    non-diagonal couplings) for a variety of fits to COHERENT data as
    labeled in the figure. In all cases shown in the left panel, the
    QF, nuclear FF and background assumptions are those employed in
    the data release~\recite{Akimov:2018vzs}.  On the right panel we
    show the dependence on the assumptions for steady background
    modeling, nuclear FF, and QF. For simplicity, in this figure we
    set all off-diagonal NSI parameters to zero, but it should be kept
    in mind that the results of our global analysis presented in
    Sec.~\ref{sec:resglob} have been obtained allowing all operators
    simultaneously in the fit.}
  \label{fig:epscoh}
\end{figure}

In the framework of NSI, the constraints on the weak charges derived
above can be directly translated into constraints on the effective
Wilson coefficients. This is shown in Fig.~\ref{fig:epscoh}, where we
plot the allowed regions for the two relevant flavor-diagonal NSI
couplings after setting the flavor-changing ones to zero.  In doing so
we notice that, as discussed in Ref.~\cite{Esteban:2018ppq}, the fact
that the neutron/proton ratio in the two target nuclei is very similar
($N_\text{Cs} \big/ Z_\text{Cs} \simeq 1.419$ for cesium and
$N_\text{I} \big/ Z_\text{I} \simeq 1.396$ for iodine) allows to
approximate Eq.~\eqref{eq:Qalpha-nsi} as:
\begin{equation}
  \label{eq:Qalpha-coh}
  \Qwq_\alpha (\vec\Eps) \propto
  \big[ (g_p^V + Y_n^\text{coh} g_n^V)
    + \Eps_{\alpha\alpha}^\text{coh} \big]^2 
  + \sum_{\beta\neq\alpha}
  \big( \Eps_{\alpha\beta}^\text{coh} \big)^2
\end{equation}
with an average value $Y_n^\text{coh} = 1.407$ and
\begin{equation}
  \label{eq:eps-nucleon}
  \Eps_{\alpha\beta}^\text{coh}
  \equiv \Eps_{\alpha\beta}^p + Y_n^\text{coh} \Eps_{\alpha\beta}^n \,,
  \qquad
  \Eps_{\alpha\beta}^p \equiv 2\Eps_{\alpha\beta}^u + \Eps_{\alpha\beta}^d \,,
  \qquad
  \Eps_{\alpha\beta}^n \equiv 2\Eps_{\alpha\beta}^d + \Eps_{\alpha\beta}^u \,.
\end{equation}
From Eq.~\eqref{eq:Qalpha-nsi} it is evident that COHERENT can only be
sensitive to a certain combination of NSI operators
$\Eps_{\alpha\beta}^\text{coh}$, which are ultimately determined by
just two factors: (a) the value of $Y_n^\text{coh}$, which depends on
the nuclei in the detector, and (b) the strength of the coupling of
the new interaction to up and down quarks (or, equivalently, to
protons and neutrons). In fact, using the $\eta$ parametrization in
Eq.~\eqref{eq:eps-fact}, $\Eps_{\alpha\beta}^\text{coh}$ can be
written as:
\begin{equation}
  \label{eq:eps-coh}
  \Eps_{\alpha\beta}^\text{coh}
  = \sqrt{5} \left( \cos\eta + Y_n^\text{coh} \sin\eta \right)
  \Eps_{\alpha\beta}^\eta \,.
\end{equation}
It is clear from the expressions above that the best-fit value and
allowed ranges of $\Eps_{\alpha\beta}^\text{coh}$ implied by COHERENT
are independent of $\eta$. Once these have been determined, the
corresponding bounds on the associated couplings
$\Eps_{\alpha\beta}^\eta$ for a given NSI model (identified by a
particular value of $\eta$) can be obtained in a very simple way, by
just rescaling the values of $\Eps_{\alpha\beta}^\text{coh}$ as
$[\sqrt{5} (\cos\eta + Y_n^\text{coh} \sin\eta)]^{-1}$.  For example,
the results in Fig.~\ref{fig:epscoh} can be immediately translated in
the corresponding ranges for NSI models where the new interaction
couples only to $f=u$, $f=d$, or $f=p$ ($\eta \approx 26.6^\circ$,
$63.4^\circ$, and $0$, respectively), after rescaling the bounds on
$\Eps_{\alpha\beta}^\text{coh}$ by the corresponding factors of
$0.293$, $0.262$, and $1$ in each case. Furthermore, from
Eq.~\eqref{eq:eps-coh} it becomes evident that, for NSI models with
$\eta = \arctan(-1/Y_n^\text{coh}) \approx -35.4^\circ$, no bound can
be derived from COHERENT data.

The impact of the timing information on the fit can be readily
observed from the left panel in Fig.~\ref{fig:epscoh}.  In the absence
of any timing information and using total rate information alone, it
is straightforward to show that, if the experiment observes a result
compatible with the SM expectation, the allowed confidence regions in
this plane should obey the equation of an ellipse. This automatically
follows from Eqs.~\eqref{eq:Qbands} and~\eqref{eq:Qalpha-coh}:
\begin{equation}
  \label{eq:ellipse}
   \frac{1}{3} [R + \Eps_{ee}^\text{coh}]^2
  + \frac{2}{3} [R + \Eps_{\mu\mu}^\text{coh}]^2
  = R^2 \,,
\end{equation}
where $R \equiv g_p^V + Y_n^\text{coh} g_n^V \approx -0.68$. This is
also shown by our numerical results in the left panel of
Fig.~\ref{fig:epscoh} which do not include timing information in the
fit (dashed and dot-dashed contours).

While the inclusion of a non-zero $\Eps_{ee}^\text{coh}$ can be
compensated by a change in $\Eps_{\mu\mu}^\text{coh}$ that brings the
total number of events in the opposite direction without significantly
affecting the delayed events, this would be noticed in the prompt
event distribution once timing information is added to the fit. In
particular, too large/small values of $\Eps_{ee}^\text{coh}$ would
require a consequent modification of the $\Eps_{\mu\mu}^\text{coh}$ to
recover the same event rate in the delayed time bins, which is however
not allowed by the prompt events observed. Thus, once timing
information is included in the fit the ellipse is broken in this plane
and two separate minima are obtained (dotted and solid lines).

It should also be noted that the central region in
Fig.~\ref{fig:epscoh} (around the centre of the ellipse in
Eq.~\eqref{eq:ellipse}, $\Eps_{ee}^\text{coh} =
\Eps_{\mu\mu}^\text{coh} = -R$) can be excluded at COHERENT
\emph{only} in the case when the off-diagonal NSI operators are not
included in the fit. This is so because in this region the effect of
the diagonal parameters leads to a destructive interference in the
total cross section and therefore to a reduction of the number of
events, in contrast with the experimental observation. Once the
off-diagonal operators are introduced this is no longer the case and
the central region becomes allowed~\cite{Giunti:2019xpr}. However,
since global neutrino oscillation data provide tight constraints on
the off-diagonal NSI operators, in our results the two minima remain
separate even after the off-diagonal operators are allowed in the fit,
as we will show in Sec.~\ref{sec:resglob}.

\subsection{Global analysis of COHERENT and oscillation data}
\label{sec:resglob}

We now present the results of the global analysis of oscillation plus
COHERENT data. To this end we construct a combined $\chi^2$ function
\begin{equation}
  \chi^2_\text{global}(\vec\Eps) = \min\limits_{\vec\omega} \left[
    \chi^2_\text{OSC}(\vec\omega, \vec\Eps)
    + \chi^2_\text{COH}(\vec\Eps) \right] \,,
\end{equation}
where we denote by $\vec\omega \equiv \{ \theta_{ij},
\delta_\text{CP}, \Dmq_{ji} \}$ the ``standard'' $3\nu$ oscillation
parameters.  For the detailed description of methodology and data
included in $\chi^2_\text{OSC}$ we refer to the comprehensive global
fit in Ref.~\cite{Esteban:2018ppq} performed in the framework of
three-flavor oscillations plus NSI with quarks parametrized as
Eq.~\eqref{eq:eps-fact}. In this work we minimally update the results
from Ref.~\cite{Esteban:2018ppq} to account for the latest LBL data
samples included in NuFIT-4.1~\cite{nufit-4.1}. To keep the fit
manageable in Ref.~\cite{Esteban:2018ppq} only the CP-conserving case
with real NSI and $\delta_\text{CP} \in \{ 0, \pi \}$ was considered,
and consequently the T2K and NO$\nu$A appearance data (which exhibit
substantial dependence on the leptonic CP phase) were not included in
the fit. Here we follow the same approach and consistently update only
the disappearance samples from these experiments.\footnote{For a
  discussion of CP violation in the presence of NSI see
  Ref.~\cite{Esteban:2019lfo}.}

\begin{figure}\centering
  \includegraphics[width=0.9\textwidth]{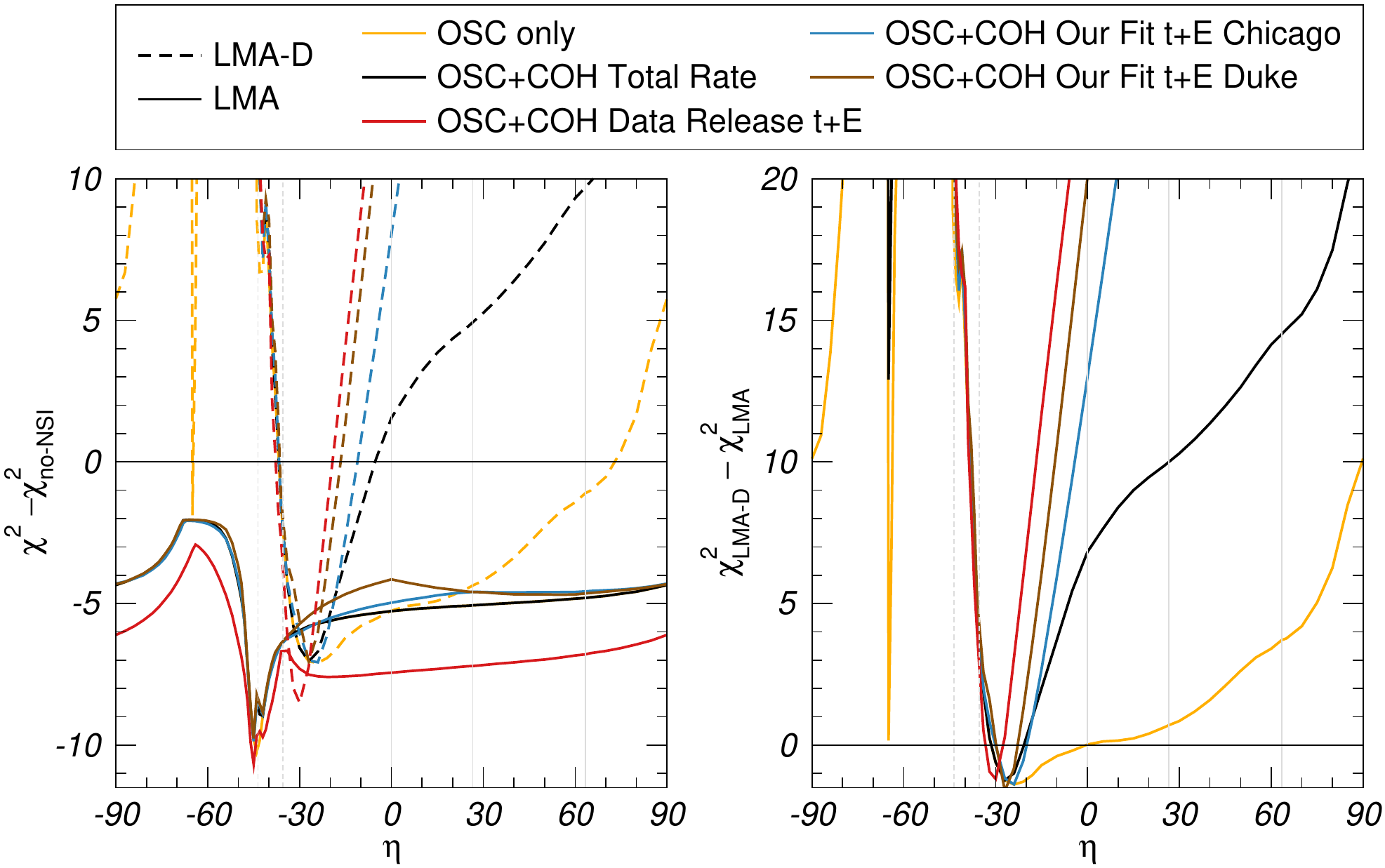}
  \caption{Left: $\chi^2_\text{LMA}(\eta) - \chi^2_\text{no-NSI}$
    (full lines) and $\chi^2_\text{LMA-D}(\eta) -
    \chi^2_\text{no-NSI}$ (dashed lines) for the analysis of different
    data combinations (as labeled in the figure) as a function of the
    NSI quark coupling parameter $\eta$.  Right:
    $\chi^2_\text{dark}-\chi^2_\text{light}\equiv
    \chi^2_\text{LMA-D}(\eta) - \chi^2_\text{LMA}(\eta)$ as a function
    of $\eta$. See text for details.}
  \label{fig:chisq-eta}
\end{figure}

Figure~\ref{fig:chisq-eta} shows the impact of COHERENT on the global
fit (left panel) as well as on the LMA-D degeneracy (right panel).  In
doing so, we have defined the functions $\chi^2_\text{LMA}(\eta)$ and
$\chi^2_\text{LMA-D}(\eta)$, obtained by marginalizing
$\chi^2_\text{global}(\vec\omega, \vec\Eps)$ over both $\vec\omega$
and $\vec\Eps$ for a given value of $\eta$, with the constraint
$\theta_{12} < 45^\circ$ (in the LMA case) and $\theta_{12} >
45^\circ$ (for the LMA-D).  With these definitions, we show in the
left panel the differences $\chi^2_\text{LMA}(\eta) -
\chi^2_\text{no-NSI}$ (full lines) and $\chi^2_\text{LMA-D}(\eta) -
\chi^2_\text{no-NSI}$ (dashed lines), where $\chi^2_\text{no-NSI}$ is
the minimum $\chi^2$ for standard $3\nu$ oscillations (\textit{i.e.},
setting all the NSI parameters to zero). Then, in the right panel we
show the values of $\chi^2_\text{LMA-D}(\eta) -
\chi^2_\text{LMA}(\eta)$, which quantifies the relative quality of the
LMA and LMA-D solutions as a function of $\eta$.

First, from the left panel in Fig.~\ref{fig:chisq-eta} we notice that
the introduction of NSI leads to a substantial improvement of the fit
already for the LMA solution (solid lines) with respect to the
oscillation data analysis, resulting in a sizable decrease of the
minimum $\chi^2_\text{LMA}$ with respect to the standard oscillation
scenario.  This is mainly driven by a well-known tension (although
mild, at the level of $\Delta\chi^2\sim 7.4$ in the present analysis)
between solar and KamLAND data in the determination of $\Dmq_{21}$.
As seen in the figure, the inclusion of NSI improves the combined fit
by about $2.2\sigma$ over a broad range of values of $\eta$.  The
improvement is maximized for NSI models with values of $\eta$ for
which the effect is largest in the Sun without entering in conflict
with terrestrial experiments.  This occurs for $\eta \simeq -44^\circ$
(as for this value the NSI in the Earth matter essentially cancel) and
leads to an improvement of about $10$ units in $\chi^2$
(\textit{i.e.}, a $\sim 3.2\sigma$ effect).  From the figure we also
conclude that adding the information from COHERENT on rate only, as
well as on timing and energy (t+E), still allows for this improved fit
in the LMA solution for most values of $\eta$. Indeed, the maximum
effect at $\eta \simeq -44^\circ$ still holds after the combination
since it falls very close to $-35.4^\circ$, for which NSI effects
cancel at COHERENT as seen in Eq.~\eqref{eq:eps-coh}.  Interestingly,
the improvement is slightly larger for the combination with COHERENT
t+E data using the data release assumptions. This is so because, as
described in the previous section, in this case the fit pulls the weak
charge $\Qwq_e$ towards zero (see Fig.~\ref{fig:qw}) while leaving the
value of $\Qwq_\mu$ around the SM expectation. Such situation can be
easily accommodated by invoking diagonal NSI operators and, in
particular, favors the non-standard values $\Eps_{ee}^\eta -
\Eps_{\mu\mu}^\eta \neq 0$, thus bringing the fit to a better
agreement with solar+KamLAND oscillation data.

Most importantly, Fig.~\ref{fig:chisq-eta} shows that the main impact
of including COHERENT data in the analysis is on the status of the
LMA-D degeneracy. We see in the figure that with oscillation data
alone the LMA-D solution is still allowed at $3\sigma$ for a wide
range of NSI models ($-38^\circ \lesssim \eta \lesssim 87^\circ$, as
well as a narrow window around $\eta \simeq -65^\circ$) and, in fact,
for $-31^\circ \lesssim \eta\lesssim 0^\circ$ it provides a slightly
better global fit than the LMA solution. The addition of COHERENT to
the analysis of oscillation data disfavors the LMA-D degeneracy for
most values of $\eta$, and the inclusion of the timing and energy
information makes this conclusion more robust.  More quantitatively we
find that, when COHERENT results are taken into account, the LMA-D is
allowed below $3\sigma$ only for values of $\eta$ in the following
ranges:
\begin{equation}
  \label{eq:LMADcoh}
  \begin{aligned}
    -38^\circ &\lesssim \eta \lesssim \hphantom{+}15^\circ
    &&\text{COHERENT Total Rate,}
    \\
    -38^\circ &\lesssim \eta \lesssim -18^\circ
    && \text{COHERENT t+E Data Release,}
    \\
    -38^\circ &\lesssim \eta \lesssim \hphantom{0}{-6^\circ}
    && \text{COHERENT t+E Our Fit Chicago,}
    \\
    -38^\circ &\lesssim \eta \lesssim -12^\circ
    &&\text{COHERENT t+E Our Fit Duke.}
  \end{aligned}
\end{equation}

\begin{figure}\centering
  \includegraphics[width=\textwidth]{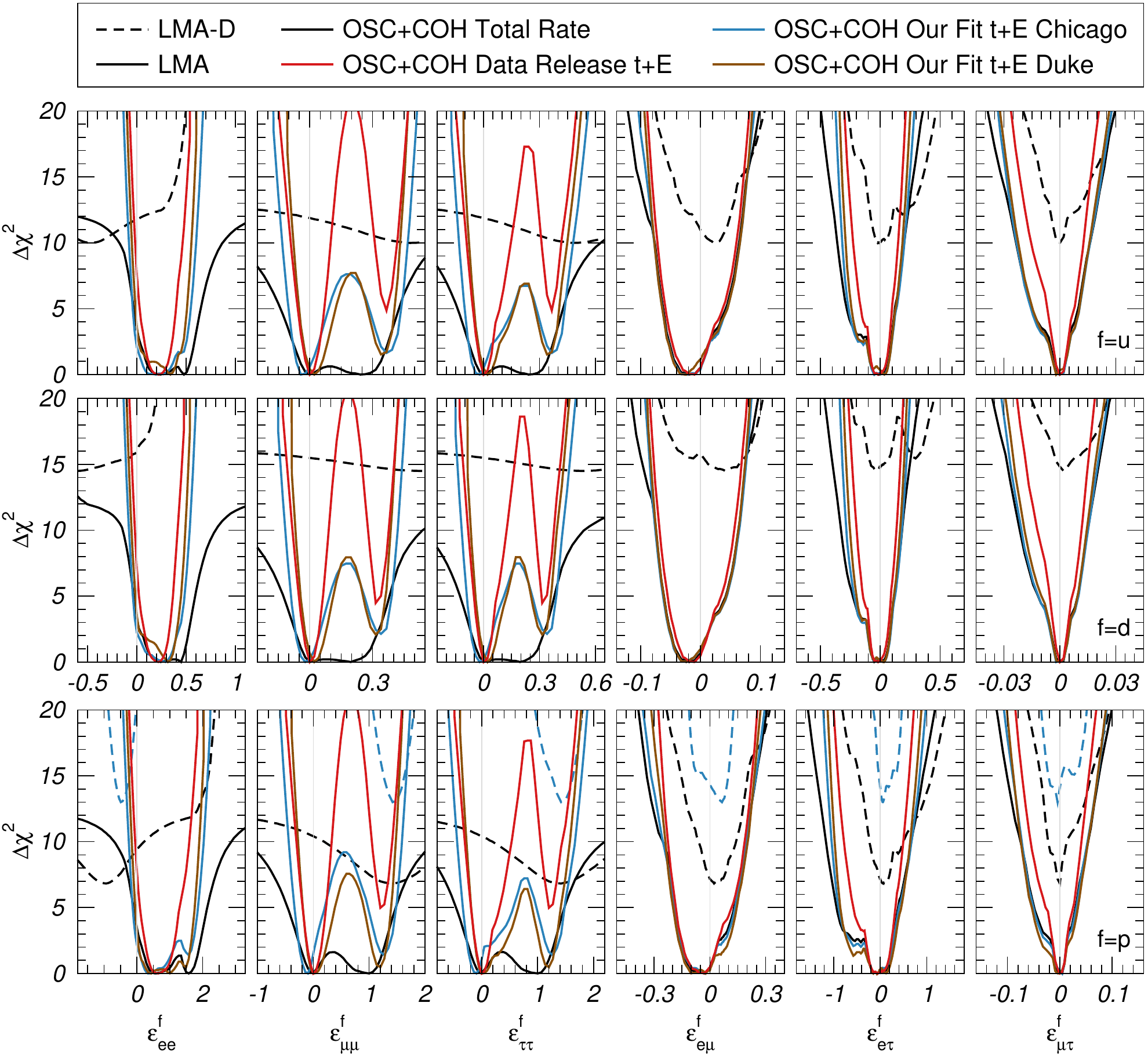}
  \caption{Dependence of the $\Delta\chi^2_\text{global}$ function on
    the NSI couplings with up quarks (upper row), down quark (central
    row) and, protons (lower row) for the global analysis of
    oscillation and COHERENT data. In each panel
    $\chi^2_\text{global}$ is marginalized with respect to the other
    five NSI couplings not shown and with respect to the oscillation
    parameters for the LMA (solid) and LMA-D (dashed) solutions.  The
    different curves correspond to the different variants of the
    COHERENT analysis implemented in this work: total rate (black),
    t+E Data Release (red), t+E with QF-C (blue), and t+E with QF-D
    (brown); see text for details.}
  \label{fig:chisq-qrk}
\end{figure}

Finally, we provide in Fig.~\ref{fig:chisq-qrk} the $\chi^2$ profiles
for each of the six NSI coefficients after marginalization over the
undisplayed oscillation parameters and the other five NSI coefficients
not shown in a given panel. We show these results for three
representative cases of NSI models including couplings to up quarks
only, down quarks only and to protons. The corresponding $2\sigma$
ranges are also provided in Tab.~\ref{tab:ranges} for convenience.
This figure shows that the LMA-D solution for NSI models that couple
only to protons ($\eta = 0$) can only be excluded beyond $3\sigma$ if
both energy and timing information are included for COHERENT, in
agreement with Eq.~\eqref{eq:LMADcoh}. From this figure we also see
that for the LMA solution the allowed ranges for the off-diagonal NSI
couplings are only moderately reduced by the addition of the COHERENT
results and, moreover, the impact of the energy and timing information
is small in these cases.  This is because they are already very well
constrained by oscillation data alone.

\begin{sidewaystable}
  \centering
  \begin{tabular}{|l|c|c|c|c|}
    \hline
    & Total Rate & Data Release t+E & Our Fit t+E Chicago & Our Fit t+E Duke \\
    \hline
    $\Eps_{ee}^u$
    & $[-0.012, +0.621]$
    & $[+0.043, +0.384]$
    & $[-0.032, +0.533]$\hfill~
    & $[-0.004, +0.496]$\hfill~
    \\
    $\Eps_{\mu\mu}^u$
    & $[-0.115, +0.405]$
    & $[-0.050, +0.062]$
    & $[-0.094, +0.071] \oplus [+0.302, +0.429]$
    & $[-0.045, +0.108] \oplus [+0.290, +0.399]$
    \\
    $\Eps_{\tau\tau}^u$
    & $[-0.116, +0.406]$
    & $[-0.050, +0.065]$
    & $[-0.095, +0.125] \oplus [+0.302, +0.428]$
    & $[-0.045, +0.141] \oplus [+0.290, +0.399]$
    \\
    $\Eps_{e\mu}^u$
    & $[-0.059, +0.033]$
    & $[-0.055, +0.027]$
    & $[-0.060, +0.036]$\hfill~
    & $[-0.060, +0.034]$\hfill~
    \\
    $\Eps_{e\tau}^u$
    & $[-0.250, +0.110]$
    & $[-0.141, +0.090]$
    & $[-0.243, +0.118]$\hfill~
    & $[-0.222, +0.113]$\hfill~
    \\
    $\Eps_{\mu\tau}^u$
    & $[-0.012, +0.008]$
    & $[-0.006, +0.006]$
    & $[-0.013, +0.009]$\hfill~
    & $[-0.012, +0.009]$\hfill~
    \\
    \hline
    $\Eps_{ee}^d$
    & $[-0.015, +0.566]$
    & $[+0.036, +0.354]$
    & $[-0.030, +0.468]$\hfill~
    & $[-0.006, +0.434]$\hfill~
    \\
    $\Eps_{\mu\mu}^d$
    & $[-0.104, +0.363]$
    & $[-0.046, +0.057]$
    & $[-0.083, +0.077] \oplus [+0.278, +0.384]$
    & $[-0.037, +0.099] \oplus [+0.267, +0.356]$
    \\
    $\Eps_{\tau\tau}^d$
    & $[-0.104, +0.363]$
    & $[-0.046, +0.059]$
    & $[-0.083, +0.083] \oplus [+0.279, +0.383]$
    & $[-0.038, +0.104] \oplus [+0.268, +0.354]$
    \\
    $\Eps_{e\mu}^d$
    & $[-0.058, +0.032]$
    & $[-0.052, +0.024]$
    & $[-0.059, +0.034]$\hfill~
    & $[-0.058, +0.034]$\hfill~
    \\
    $\Eps_{e\tau}^d$
    & $[-0.198, +0.103]$
    & $[-0.106, +0.082]$
    & $[-0.196, +0.107]$\hfill~
    & $[-0.181, +0.101]$\hfill~
    \\
    $\Eps_{\mu\tau}^d$
    & $[-0.008, +0.008]$
    & $[-0.005, +0.005]$
    & $[-0.008, +0.008]$\hfill~
    & $[-0.007, +0.008]$\hfill~
    \\
    \hline
    $\Eps_{ee}^p$
    & $[-0.035, +2.056]$
    & $[+0.142, +1.239]$
    & $[-0.095, +1.812]$\hfill~
    & $[-0.024, +1.723]$\hfill~
    \\
    $\Eps_{\mu\mu}^p$
    & $[-0.379, +1.402]$
    & $[-0.166, +0.204]$
    & $[-0.312, +0.138] \oplus [+1.036, +1.456]$
    & $[-0.166, +0.337] \oplus [+0.952, +1.374]$
    \\
    $\Eps_{\tau\tau}^p$
    & $[-0.379, +1.409]$
    & $[-0.168, +0.257]$
    & $[-0.313, +0.478] \oplus [+1.038, +1.453]$
    & $[-0.167, +0.582] \oplus [+0.950, +1.382]$
    \\
    $\Eps_{e\mu}^p$
    & $[-0.179, +0.112]$
    & $[-0.174, +0.086]$
    & $[-0.179, +0.120]$\hfill~
    & $[-0.187, +0.131]$\hfill~
    \\
    $\Eps_{e\tau}^p$
    & $[-0.877, +0.340]$
    & $[-0.503, +0.295]$
    & $[-0.841, +0.355]$\hfill~
    & $[-0.817, +0.386]$\hfill~
    \\
    $\Eps_{\mu\tau}^p$
    & $[-0.041, +0.025]$
    & $[-0.020, +0.019]$
    & $[-0.044, +0.026]$\hfill~
    & $[-0.048, +0.030]$\hfill~
    \\
    \hline
  \end{tabular}
  \caption{$2\sigma$ allowed ranges for the NSI couplings
    $\Eps_{\alpha\beta}^u$, $\Eps_{\alpha\beta}^d$ and
    $\Eps_{\alpha\beta}^p$ as obtained from the global analysis of
    oscillation plus COHERENT data. See text for details.}
  \label{tab:ranges}
  \afterpage\clearpage
\end{sidewaystable}

More interestingly, the addition of COHERENT data allows to derive
constraints on each of the diagonal parameters separately and, for
those, the timing (and to less degree energy) information has a
quantitative impact. In particular we see in the figure the appearance
of the two minima corresponding to the degenerate solutions for
$\Eps_{\mu\mu}^\text{coh}$ in Fig.~\ref{fig:epscoh}, obtained after
the inclusion of timing information for COHERENT. Noticeably, now the
non-standard solution (obtained for $\Eps_{\mu\mu}^f \neq 0$) is
partially lifted by the combination with oscillation data, but it
remains well allowed around $\sim 2\sigma$ depending on the
assumptions for the COHERENT analysis.  Fig.~\ref{fig:chisq-qrk} also
shows the corresponding two minima for $\Eps_{\tau\tau}^f$ arising
from the combination of the information on $\Eps_{\tau\tau}^f -
\Eps_{\mu\mu}^f$ and $\Eps_{ee}^f - \Eps_{\mu\mu}^f$ from the
oscillation experiments with the constraints on $\Eps_{\mu\mu}^f$ from
the COHERENT t+E data.  In particular, in all the three cases $f=u$,
$d$, $p$ shown in the figure the bound on $\Eps_{\tau\tau}^f$ becomes
about two orders of magnitude stronger than previous indirect
(loop-induced) limits~\cite{Davidson:2003ha} when the t+E analysis
with the data release assumptions is used. Indeed this conclusion
holds for most $\eta$ values, with exception of $\eta \sim -45^\circ$
to $-35^\circ$ for which NSI effects are suppressed in either the
Earth matter or in COHERENT. In particular, for $\eta=-35.4^\circ$ the
NSI effects in COHERENT totally cancel, as described above, and
consequently no separate determination of the three diagonal
parameters is possible around such value.

\section{Summary and Conclusions}
\label{sec:summary}

From a completely model-independent approach, a useful way to
parametrize the effects of new physics models on low-energy
experiments is through the inclusion of additional higher-dimensional
operators to the Standard Model Lagrangian. In this context,
four-fermion operators leading to neutral-current interactions between
neutrinos and quarks, usually referred to as neutrino non-standard
interactions (NSI), can lead to novel effects in both neutrino
propagation, production and detection processes.  Although models
leading to large NSI in the neutrino sector are challenging to build
from a high-energy theory, new physics models invoking light mediators
are able to evade the tight constraints in the charged lepton sector,
and can lead to observable consequences in neutrino experiments. While
charged-current NSI are better constrained by precision measurements
of muon and meson decays, neutral-current NSI are harder to constrain
experimentally and, in fact, the best bounds in this case come from
global fits to neutrino data.

In this work, we have combined neutrino oscillation data with the
latest results obtained for coherent neutrino-nucleus scattering data
at the COHERENT experiment, which provide both energy and timing
information. The results of our analysis are used to constrain the
Wilson coefficients of the whole set of neutral-current operators
leading to NSI involving up and down quarks simultaneously. Therefore,
our conclusions extend to a general class of new physics models with
arbitrary ratios (parametrized by an angle $\eta$) between the
strength of the operators for up and down quarks, and mediated by $Z'$
with masses above $M_\text{med} \sim \mathcal{O}(10)$~MeV.

We have quantified the dependence of our results for COHERENT with
respect to the choice of quenching factor, nuclear form factor, and
the treatment of the backgrounds. We find that the implementation of
the steady-state background has a strong impact on the results of the
analysis of COHERENT due to a slight background excess in the first
two bins, which is present in both the coincident and anti-coincident
data samples provided by the collaboration. Once this effect has been
accounted for in the modeling of the expected backgrounds, the choice
of quenching factor and nuclear form factor has a minor impact on the
results obtained from the fit.

We find that the inclusion of COHERENT timing information affects the
global fit significantly and, most notably, has a large impact on the
constraints that can be derived for the flavor-diagonal NSI operators,
for which separate constraints can only be derived after combination
of COHERENT and oscillation data.

Furthermore, the presence of NSI is known to introduce a degeneracy in
the oscillation probabilities for neutrinos propagating in matter,
leading in particular to the appearance of the LMA-Dark (LMA-D)
solution.  We find that the inclusion of COHERENT to the analysis of
oscillation data disfavors the LMA-D degeneracy for NSI models over a
wide range of $\eta$, and the addition of the timing and energy
information makes this conclusion more robust, see
Eq.~\eqref{eq:LMADcoh} and Fig.~\ref{fig:chisq-eta}. In particular,
the LMA-D solution for NSI models that couple only to protons ($\eta =
0$) can only be excluded beyond $3\sigma$ once both energy and timing
information are included for the COHERENT data.

Finally, the introduction of NSI is known to alleviate the well-known
(albeit mild) tension between solar and KamLAND data in the
determination of the $\Dmq_{21}$, thus leading to an overall
improvement in the quality of the global fit to oscillation data.  We
find that this still remains the case after the inclusion of COHERENT
results.

\section*{Acknowledgments}

The authors warmly thank Javier Menéndez for providing nuclear form
factors for CsI, and Juan Collar for providing the QF data from
Ref.~\cite{Collar:2019ihs}. They are also grateful to Diego
Aristizabal, Phil Barbeau, Bashkar Dutta, Carlo Giunti and Martin
Hoferichter for useful discussions.  MCGG thanks the Department of
Physics at Columbia University for their hospitality. PC thanks the
CERN Theory Division and the Fermilab Theory Group for their support
and hospitality during the final stages of this work. This work was
supported by the MINECO grant FPA2016-76005-C2-1-P, by the MINECO
FEDER/UE grants FPA2015-65929-P, FPA2016-78645-P and FPA2017-85985-P,
by PROMETEO/2019/083, by USA-NSF grants PHY-1620628, by EU Networks
FP10 ITN ELUSIVES (H2020-MSCA-ITN-2015-674896) and INVISIBLES-PLUS
(H2020-MSCA-RISE-2015-690575), by the ``Severo Ochoa'' program grant
SEV-2016-0597 of IFT and by AGAUR (Generalitat de Catalunya) grant
2017-SGR-929.  IE acknowledges support from the FPU program fellowship
FPU15/0369.

\newpage
\section*{Addendum}
In this addendum we re-assess the constraints on Non-Standard Interactions
(NSI) from the combined analysis of data from oscillation experiments
and from COHERENT after including the  new data released since the
publication of this work~\cite{Coloma:2019mbs},
in particular those presented at
the Neutrino2020 conference. New data considered includes
the latest total energy spectrum and the day-night asymmetry
of the SK4 2970-day sample presented at
Neutrino2020~\cite{SK:nu2020}, and the latest results from long-baseline (LBL)
experiments  T2K~\cite{Abe:2019vii, T2K:nu2020} and NOvA~\cite{Acero:2019ksn,NOvA:nu2020}. In addition, we have updated the reactor experiments
Double-Chooz~\cite{DoubleChooz:2019qbj,DoubleC:nu2020} to 1276/587 days of
far/near detector data and RENO~\cite{Bak:2018ydk, RENO:nu2020}  to 2908
days of exposure.

\begin{figure}\centering
  \includegraphics[width=0.9\textwidth]{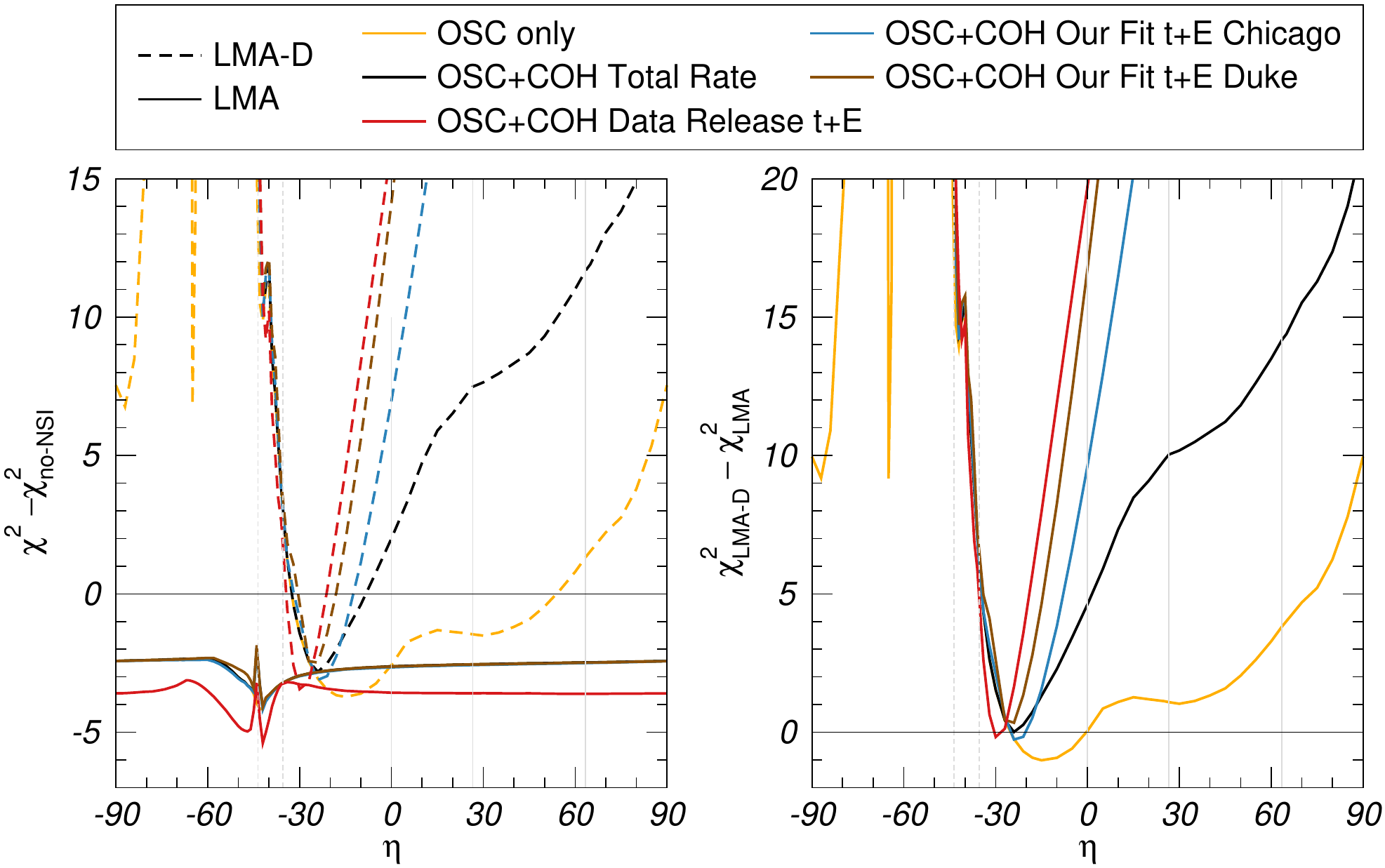}
  \caption{Left: $\chi^2_\text{LMA}(\eta) - \chi^2_\text{no-NSI}$
    (full lines) and $\chi^2_\text{LMA-D}(\eta) -
    \chi^2_\text{no-NSI}$ (dashed lines) for the analysis of different
    data combinations (as labeled in the figure) as a function of the
    NSI quark coupling parameter $\eta$. All solid lines but the red one
   falls on top of each other.  Right:
    $\chi^2_\text{dark}-\chi^2_\text{light}\equiv
    \chi^2_\text{LMA-D}(\eta) - \chi^2_\text{LMA}(\eta)$ as a function
    of $\eta$. See text for details.}
  \label{fig:chisq-eta-20}
\end{figure}

\begin{figure}\centering
  \includegraphics[width=\textwidth]{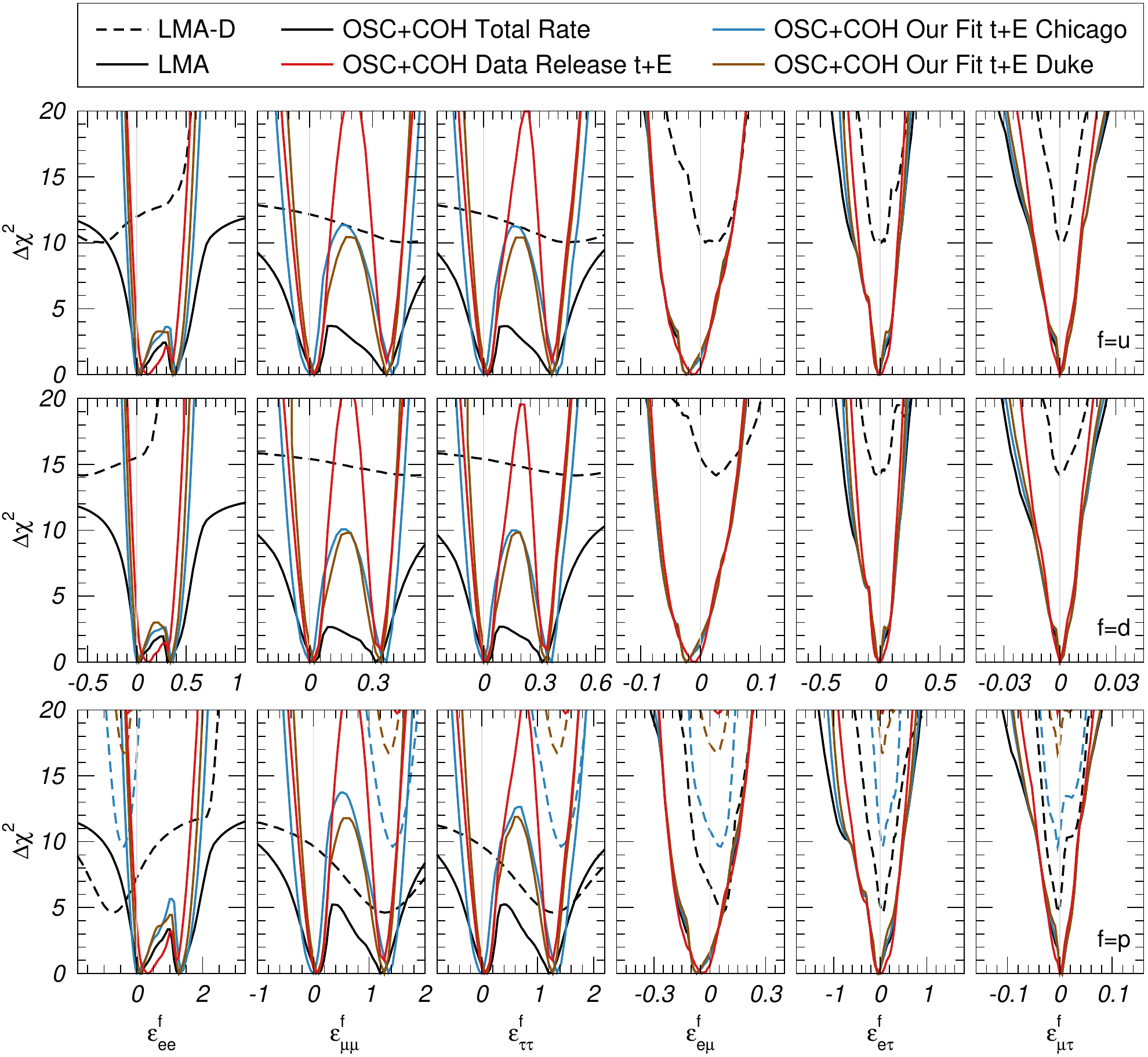}
  \caption{Dependence of the $\Delta\chi^2_\text{global}$ function on
    the NSI couplings with up quarks (upper row), down quark (central
    row) and, protons (lower row) for the global analysis of
    oscillation and COHERENT data. In each panel
    $\chi^2_\text{global}$ is marginalized with respect to the other
    five NSI couplings not shown and with respect to the oscillation
    parameters for the LMA (solid) and LMA-D (dashed) solutions.  The
    different curves correspond to the different variants of the
    COHERENT analysis implemented in this work: total rate (black),
    t+E Data Release (red), t+E with QF-C (blue), and t+E with QF-D
    (brown); see text for details.}
  \label{fig:chisq-qrk-20}
\end{figure}

The main effect driven by the new results concerns the analysis of
solar oscillation data. The  quantification of the effects in the
oscillation analysis has been presented in a separate Addendum to
Ref.~\cite{Esteban:2018ppq}. Here we quantify the induced changes
in the results of the combined analysis of oscillation data with
COHERENT results, which were contained in Sec.~\ref{sec:resglob}. In particular we present in Figs.~\ref{fig:chisq-eta-20}
and~\ref{fig:chisq-qrk-20} the new version of Figs.~\ref{fig:chisq-eta} and \ref{fig:chisq-qrk}, and
in Table~\ref{tab:ranges1-20} the new version of Table~\ref{tab:ranges}.

In brief, in the left panel in  Fig.~\ref{fig:chisq-eta} we found
that the introduction of NSI lead to a substantial improvement of the fit
already for the LMA solution (solid lines) with respect to the
oscillation data analysis, resulting in a sizable decrease of the
minimum $\chi^2_\text{LMA}$ with respect to the standard oscillation
scenario.  This was driven by a well-known tension
at the level of $\Delta\chi^2\sim 7.4$  
between solar and KamLAND data in the determination of $\Dmq_{21}$.
Correspondingly the inclusion of NSI improved the combined fit
by about $2.2\sigma$ over a broad range of values of $\eta$.
As discussed in Ref.~\cite{Esteban:2020cvm}, 
with the updated SK4 solar data the tension between the best
fit $\Dmq_{21}$ of KamLAND and that of the solar results has
decreased to  $\Delta\chi^2_\text{solar} = 1.3$. 
So now in the left panel in Fig.~\ref{fig:chisq-eta-20} we see that
for the LMA solution the combined global fit with NSI leads to a decrease
of about 2 units in $\chi^2$ for most values of $\eta$ and for most
variants of the COH analysis. The only exception is the analysis  
of the combination with COHERENT t+E data using the data release assumptions,
for which including NSI can improve the fit in LMA by $\Delta \chi^2 \sim 4$
for most values of $\eta$.

Concerning the status of the LMA-D degeneracy, we find that when
COHERENT total rate results are taken into account and we include the
new oscillation data, LMA-D is allowed below $3\sigma$ with respect to
LMA for a slightly wider range of values of $\eta$. This is a
consequence of the increase of $\chi^2_\text{LMA}$. Quantitatively,
LMA-D is now allowed at 3$\sigma$ for values of $\eta$ in the
following ranges:
\begin{equation}
  \label{eq:LMADcoh-20}
  \begin{aligned}
    -37^\circ &\lesssim \eta \lesssim \hphantom{+}20^\circ
    &&\text{COHERENT Total Rate,}
    \\
    -37^\circ &\lesssim \eta \lesssim -14^\circ
    && \text{COHERENT t+E Data Release,}
    \\
    -37^\circ &\lesssim \eta \lesssim \hphantom{0}{0^\circ}
    && \text{COHERENT t+E Our Fit Chicago,}
    \\
    -37^\circ &\lesssim \eta \lesssim -9^\circ
    &&\text{COHERENT t+E Our Fit Duke.}
  \end{aligned}
\end{equation}

Figure.~\ref{fig:chisq-qrk-20} contains the updated $\Delta \chi^2$ profiles
for each of the six NSI coefficients after marginalization over the
undisplayed oscillation parameters and the other five NSI coefficients
not shown in a given panel, for three
representative cases of NSI models including couplings to up quarks
only, down quarks only and to protons. The corresponding $2\sigma$
ranges are also provided in Tab.~\ref{tab:ranges1-20} for convenience.
The main difference introduced by the new oscillation data is that 
now the two minima corresponding to the degeneracy on  
$\Eps_{\mu\mu}^\text{coh}$  obtained after the inclusion of timing information
for COHERENT, is no longer broken after combination with the updated
oscillation data.

This leads to the appearance of disconnected allowed ranges at 2$\sigma$
when comparing Table~\ref{tab:ranges1-20} with Table~\ref{tab:ranges}.

\begin{sidewaystable}\small
  \centering
  \begin{tabular}{|l|l|l|l|l|}
    \hline
    & \multicolumn{1}{c|}{Total Rate}
    & \multicolumn{1}{c|}{Data Release t+E}
    & \multicolumn{1}{c|}{Our Fit t+E Chicago}
    & \multicolumn{1}{c|}{Our Fit t+E Duke}
    \\
    \hline
    $\Eps_{ee}^u$
    & $[-0.067, +0.547]$
    & $[-0.004, +0.412]$
    & $[-0.059, +0.505]$
    & $[-0.031, +0.476]$
    \\
    $\Eps_{\mu\mu}^u$
    & $[-0.076, +0.455]$
    & $[-0.041, +0.067] \oplus [+0.333, +0.405]$
    & $[-0.071, +0.045] \oplus [+0.330, +0.448]$
    & $[-0.029, +0.068] \oplus [+0.309, +0.415]$
    \\
    $\Eps_{\tau\tau}^u$
    & $[-0.076, +0.455]$
    & $[-0.041, +0.067] \oplus [+0.332, +0.404]$
    & $[-0.071, +0.045] \oplus [+0.330, +0.448]$
    & $[-0.029, +0.068] \oplus [+0.309, +0.414]$
    \\
    $\Eps_{e\mu}^u$
    & $[-0.050, +0.020]$
    & $[-0.053, +0.018]$
    & $[-0.048, +0.020]$
    & $[-0.048, +0.020]$
    \\
    $\Eps_{e\tau}^u$
    & $[-0.077, +0.099]$
    & $[-0.080, +0.100]$
    & $[-0.077, +0.096]$
    & $[-0.077, +0.095]$
    \\
    $\Eps_{\mu\tau}^u$
    & $[-0.006, +0.007]$
    & $[-0.007, +0.006]$
    & $[-0.006, +0.007]$
    & $[-0.006, +0.007]$
    \\
    \hline
    $\Eps_{ee}^d$
    & $[-0.063, +0.503]$
    & $[-0.004, +0.367]$
    & $[-0.058, +0.453]$
    & $[-0.034, +0.426]$
    \\
    $\Eps_{\mu\mu}^d$
    & $[-0.072, +0.408]$
    & $[-0.038, +0.060] \oplus [+0.298, +0.366]$
    & $[-0.066, +0.043] \oplus [+0.292, +0.401]$
    & $[-0.027, +0.063] \oplus [+0.275, +0.371]$
    \\
    $\Eps_{\tau\tau}^d$
    & $[-0.072, +0.407]$
    & $[-0.038, +0.058] \oplus [+0.296, +0.365]$
    & $[-0.067, +0.042] \oplus [+0.292, +0.401]$
    & $[-0.027, +0.067] \oplus [+0.274, +0.372]$
    \\
    $\Eps_{e\mu}^d$
    & $[-0.050, +0.020]$
    & $[-0.049, +0.018]$
    & $[-0.050, +0.020]$
    & $[-0.050, +0.020]$
    \\
    $\Eps_{e\tau}^d$
    & $[-0.078, +0.098]$
    & $[-0.084, +0.094]$
    & $[-0.076, +0.098]$
    & $[-0.076, +0.097]$
    \\
    $\Eps_{\mu\tau}^d$
    & $[-0.006, +0.007]$
    & $[-0.006, +0.006]$
    & $[-0.006, +0.007]$
    & $[-0.006, +0.007]$
    \\
    \hline
    $\Eps_{ee}^p$
    & $[-0.222, +1.801]$
    & $[-0.011, +1.408]$
    & $[-0.183, +0.819] \oplus [+1.172, +1.700]$
    & $[-0.086, +0.884] \oplus [+1.083, +1.605]$
    \\
    $\Eps_{\mu\mu}^p$
    & $[-0.248, +0.282] \oplus [+0.625, +1.551]$
    & $[-0.129, +0.228] \oplus [+1.129, +1.375]$
    & $[-0.232, +0.149] \oplus [+1.135, +1.521]$
    & $[-0.097, +0.220] \oplus [+1.063, +1.410]$
    \\
    $\Eps_{\tau\tau}^p$
    & $[-0.248, +0.281] \oplus [+0.646, +1.548]$
    & $[-0.127, +0.226] \oplus [+1.125, +1.373]$
    & $[-0.232, +0.149] \oplus [+1.133, +1.519]$
    & $[-0.098, +0.221] \oplus [+1.063, +1.408]$
    \\
    $\Eps_{e\mu}^p$
    & $[-0.145, +0.058]$
    & $[-0.162, +0.053]$
    & $[-0.135, +0.058]$
    & $[-0.124, +0.058]$
    \\
    $\Eps_{e\tau}^p$
    & $[-0.239, +0.293]$
    & $[-0.233, +0.320]$
    & $[-0.237, +0.279]$
    & $[-0.239, +0.244]$
    \\
    $\Eps_{\mu\tau}^p$
    & $[-0.019, +0.021]$
    & $[-0.021, +0.017]$
    & $[-0.017, +0.021]$
    & $[-0.013, +0.021]$
    \\
    \hline
  \end{tabular}
  \caption{$2\sigma$ allowed ranges for the NSI couplings
    $\Eps_{\alpha\beta}^u$, $\Eps_{\alpha\beta}^d$ and
    $\Eps_{\alpha\beta}^p$ as obtained from the global analysis of
    oscillation plus COHERENT data. See text for details.}
  \label{tab:ranges1-20}
\end{sidewaystable}

\bibliographystyle{JHEPmod}
\bibliography{references}

\end{document}